\documentclass[twocolumn]{aastex62}

\hypersetup{linkcolor=blue,citecolor=blue,filecolor=cyan,urlcolor=blue}

\usepackage{natbib}
\usepackage{graphicx}
\usepackage{bm}
\usepackage{amsmath}
\usepackage{amsfonts}
\usepackage{amssymb}
\usepackage{xcolor}
\usepackage{mathrsfs}
\usepackage[caption=false]{subfig}
\usepackage{multirow}
\usepackage{CJK}
\usepackage{ulem}

\def\arcsec{$^{\prime\prime}$}
\def\um{$\mu \rm m$}
\def\nh3{$\rm NH_3$}
\def\Mo{$\rm M_{\odot}$}
\def\kms {km\,s$^{-1}$}

\newcommand{\skipthis}[1]{}

\def\new#1 {{\bf #1 }}

\graphicspath{{./}{figures/}}

\shortauthors{Li et al.}

\begin{document}

\title{Formation of Massive Protostellar Clusters\\
-- Observations of Massive 70 $\mu$m  Dark Molecular Clouds}

\correspondingauthor{Shanghuo Li}
\email{shanghuo.li@gmail.com, jzwang@shao.ac.cn}

\author[0000-0002-0786-7307]{Shanghuo Li}
\affil{Shanghai Astronomical Observatory, Chinese Academy of Sciences, 80 Nandan Road, Shanghai 200030, China}
\affiliation{Center for Astrophysics $|$ Harvard \& Smithsonian, 60 Garden Street, Cambridge, MA 02138, USA}
\affiliation{University of Chinese Academy of Sciences, 19A Yuquanlu, Beijing 100049, China}

\author{Qizhou Zhang}
\affiliation{Center for Astrophysics $|$ Harvard \& Smithsonian, 60 Garden Street, Cambridge, MA 02138, USA}

\author{Thushara Pillai}
\affiliation{Institute for Astrophysical Research, Boston University, 725 Commonwealth Ave, Boston, MA 02215, USA}

\author{Ian W. Stephens}
\affiliation{Center for Astrophysics $|$ Harvard \& Smithsonian, 60 Garden Street, Cambridge, MA 02138, USA}

\author{Junzhi Wang}
\affiliation{Shanghai Astronomical Observatory, Chinese Academy of Sciences, 80 Nandan Road, Shanghai 200030, China}

\author{Fei Li}
\affiliation{Shanghai Astronomical Observatory, Chinese Academy of Sciences, 80 Nandan Road, Shanghai 200030, China}
\affiliation{University of Chinese Academy of Sciences, 19A Yuquanlu, Beijing 100049, China}

\begin{abstract}
We present Submillimeter Array (SMA) observations of seven 
massive molecular clumps which are dark in the far-infrared
for wavelengths up to 70 \um. Our 1.3~mm continuum images 
reveal 44 dense cores, with gas masses ranging from 1.4 
to 77.1 \Mo. Twenty-nine dense cores have masses greater 
than 8 \Mo\ and the other fifteen dense cores have masses 
between 1.4 and 7.5 \Mo. Assuming the core density 
follows a power-law  in radius $\rho \propto r^{-b}$,  
the index $b$ is found to be between 0.6 and 2.1 with a mean 
value of 1.3. The virial analysis reveals that the dense 
cores are not in virial equilibrium.  CO outflow emission 
was detected toward  6 out of 7 molecular clumps and 
associated with 17 dense cores. For five of these cores, 
CO emissions appear to have line-wings  at velocities of 
greater than 30~km s$^{-1}$ with respect to the source 
systemic velocity, which indicates that most of the 
clumps harbor protostars and thus are not quiescent in 
star formation. The estimated outflow timescale increase 
with core mass,  which likely indicates that  massive 
cores have longer accretion timescale than that of the 
less massive ones. The fragmentation analysis shows 
that the mass of low-mass and massive cores  are 
roughly consistent with thermal and turbulent Jeans 
masses, respectively.

\end{abstract}

\keywords{Protoclusters (1297), Early-type stars (430), Star formation (1569), Massive stars (732), Protostars (1302), Stellar winds (1636), Stellar jets (1607), Radio spectroscopy (1359), Dust continuum emission (412), Submillimeter astronomy (1647)}


\section{Introduction}
\label{intrduction}
The feedback of massive stars and clusters, such as 
radiation, wind, and supernovae, accounts for most 
of the energy budget in galaxies. However, our 
understanding of massive stars and clusters formation, 
especially the early evolutionary stages is still poor 
\citep{2007ARA&A..45..481Z,2018ARA&A..56...41M}. 
The early evolutionary phases are crucial for understanding
the initial conditions of massive star and cluster 
formation. Massive ($>$10$^{2}$ \Mo) and 
dense molecular clumps ($>$10$^{3}$ cm$^{-3}$) 
that show no signs of on-going 
star formation activities are considered to be the cradle 
of the massive stars and clusters 
\citep{1998ApJ...508..721C,2006ApJ...641..389R,
2006ApJ...651L.125W,2010A&A...518L..95H,
2012A&A...547A..49R}. 
Infrared dark clouds (IRDCs) are cold and represent an 
early evolutionary stage of 
star formation molecular clump before the star have formed, 
rapidly heat, ionize, and disrupt their surrounding gas and dust 
\citep[][and references therein]{1998ApJ...508..721C,
2006ApJ...641..389R,2012A&A...547A..49R}. 
Massive IRDCs that 
exhibit little or no signs of on-going star formation 
\citep{2006ApJ...641..389R,2006ApJ...651L.125W}, 
as evidenced by the lack of powerful jets,
outflows, stellar radiation and ionization,  
provide an ideal laboratory to 
investigate the initial condition of massive star and 
cluster formation since they are not significantly 
affected by the stellar feedback

The cloud fragmentation is a crucially important process 
to form massive stars and clusters,  and it is also one 
of the main debate between two  star formation scenarios. 
The core-accretion model suggests that massive stars are 
formed in near-equilibrium status with limited 
fragmentation within massive cores 
\citep{2002Natur.416...59M,2003ApJ...585..850M}, 
while the competitive accretion model suggests that 
massive clump fragments into many low-mass cores that have
masses of around thermal Jean mass, and some of them have 
the potential to form massive stars through competitive 
accretion  \citep{2001MNRAS.323..785B,2006MNRAS.370..488B}.

Previous investigations of massive clumps reveal that 
massive cores are responsible for forming high-mass stars 
are typically an order of magnitude more massive than the 
global thermal Jeans mass of the clump and are more 
consistent with turbulent Jeans mass when the sound speed 
is replaced by the non-thermal velocity dispersion 
\citep{2009ApJ...696..268Z,2011ApJ...733...26Z,
2014MNRAS.439.3275W,2015ApJ...805..171L}. 
The properties of low-mass ($\ll$ 8 \Mo) fragments in the 
clump, on the other hand, could be interpreted as pure 
thermal Jeans fragmentation 
\citep[e.g.,][]{2014ApJ...785...42P,2015MNRAS.453.3785P,
2018ApJ...855...24P,2018A&A...617A.100B}. 
Others find that complex structures can not be 
explained by simple theoretical models 
\citep[e.g.,][]{2017MNRAS.468.3694C,2017MNRAS.464L..31H}. 
The fragmentation process of massive molecular clumps into 
dense star-forming cores remains unclear.

In addition, the majority of these fragments show signs 
of star formation activity, such as molecular outflows, 
H$_{2}$O masers, and/or class II CH$_{3}$OH masers  
\citep{2006ApJ...651L.125W,2012ApJ...745L..30W,
2015MNRAS.453.3785P,2015ApJ...804..141Z,
2017MNRAS.468.3694C,2017MNRAS.464L..31H,
2018ApJ...855...24P}. 
This is because the majority of  targets in previous 
studies are associated with infrared bright sources at 
4.5, 8 or 24 $\mu$m wavelengths. Therefore, they are not 
ideal objects for investigating the earliest evolutionary 
stage of star formation. 
The massive molecular clouds which are dark in the infrared 
for wavelengths up to 70 $\mu$m are ideal objects for 
studying the extremely early phase of star formation 
\citep{2008ApJS..179..249D,2012A&A...547A..49R,
2013ApJ...767...36S,2017MNRAS.471..100E}. 
Galactic wide surveys offer a 
unique opportunity to identify  ideal massive clouds 
that have the potential to form massive stars and clusters 
but without signs of on-going star formation. For instance, 
the ATLAGAL survey of the Galactic plane mapped over 400 
square degrees of the inner Galactic at 870 $\mu$m with 
Atacama Pathfinder Experiment (APEX) telescope
\citep{2009A&A...504..415S}.

To study the processes of massive star and cluster 
formation within massive clumps, we used the Submillimeter 
Array \citep[SMA\footnote{The Submillimeter
Array is a joint project between the Smithsonian
Astrophysical Observatory and the Academia Sinica 
Institute of Astronomy and Astrophysics, and is funded by 
the Smithsonian Institution and the Academia
Sinica.}, ][]{2004ApJ...616L...1H} to observe a 
sample of seven high-mass 70 $\mu$m dark clumps identified 
by ATLAGAL survey.  The paper is organized
as follows. First we  describe the sample selection and the
observations with the SMA in $\S$ 2.  Then, we present the 
results and analysis  in $\S$ 3.  We discuss the 
observational results in $\S$ 4, and summarize the main 
findings in $\S$ 5.

\section{Samples and  Observations}
\label{obser}
\subsection{Sample Selection}
\label{sample}
The sample of seven sources in this study, selected from 
the ATLASGAL survey of the Galactic plane 
\citep{2009A&A...504..415S}, are among the most massive 
(5 $\times$ 10$^{3}$ to 1.5 $\times$ 10$^{4}$ M$_{\odot}$) 
and the coldest 70 $\mu$m dark  ($\sim$14 K) clumps within
the distance of 6 kpc (3.5 to 5.9 kpc) from the Sun 
\citep{2014MNRAS.443.1555U}. As shown in Figure 
\ref{fig:cont1}, these sources have large reservoirs of 
gas as manifested by the presence of intense dust emission 
at 870 $\mu$m \citep{2014MNRAS.443.1555U}. These 
requirements are to ensure that the clumps likely form 
massive stars in a cluster assuming a typical star 
formation efficiency of 30\%, while being infrared dark 
at 70 \um\ is to rule out that the regions have been 
affected by significant stellar feedback 
\citep{2008ApJS..179..249D,2013ApJ...767...36S,
2017MNRAS.471..100E}.

\subsection{SMA observations}
The observations were carried out with the SMA in the 
compact configuration from July through August 2017. The 
projected baselines ranged from $\sim$8 k$\lambda$ to  
$\sim$60 k$\lambda$, where $\lambda$, around 1.3~mm, 
is the wavelength of the observations. Four targets were 
observed with seven antennas, two targets with eight 
antennas, and one target with six antennas.  The full 
width at half-maximum (FWHM) primary beam is about 
55\arcsec. The zenith opacity at 220 GHz, $\tau_{220}$, 
was between 0.04 and 0.08 during the observations. 
The typical system temperatures varied from  200 to 500 K 
depending on the source elevations and weather conditions.

Both 230 GHz and 240 GHz receivers were used in the  
observations. The simultaneous operations of two sets of 
receivers provide a total spectral bandwidth of 32 GHz, 
with the 230 GHz receiver covering  $213.4 - 221.4$ GHz 
in the lower side band (LSB) and $229.4 - 237.4$ GHz in 
the upper side band (USB), and the 240 GHz receiver 
covering $244.5 - 252.5$ GHz in the LSB and 
$260.5 - 268.5$ GHz in the USB. The SMA Wideband 
Astronomical ROACH2 Machine (SWARM) correlator 
provides a uniform channel width of 140 KHz across 
the entire spectral band. An overview of  observations 
and calibrators is summarized in Table \ref{tbl:obs}. 
The raw visibility data were averaged every two channels
before calibration using the IDL superset 
MIR\footnote{https://www.cfa.harvard.edu/$\sim$cqi/mircook.html}. 
Data from 230 GHz and 240 GHz receivers were calibrated 
separately. The Quasars 3c279 and 3c454.3 were used for
bandpass calibration. Five Quasars, 1751+096, 1743-038, 
1830+063, nrao530 and 1924+292, were employed for gain 
calibrations. Flux calibrations were performed using 
Neptune and Ganymede. 
We have applied the {\tt uti$_{-}$hayshft} on calibrated visibility 
data to fix the diurnal Doppler tracking error.

The continuum emission was generated by averaging line-free
channels over the observed spectral windows. The typical 
root mean square (rms) noise level is about  0.5 mJy 
beam$^{-1}$  in continuum images, where the typical 
synthesized beam size is about 
$\theta_{maj} \times \theta_{min}$ = 
3.4\arcsec $\times$ 2.5\arcsec. 
Only $^{12}$CO 2-1, C$^{18}$O 2-1 and SiO 5-4 lines 
were used for study in this paper.  The typical 
spectral line sensitivity is about 65 mJy beam$^{-1}$ per 
0.23 km s$^{-1}$.  The calibrated visibility data were 
exported to  Common Astronomical Software Applications 
\citep[CASA][]{2007ASPC..376..127M} for imaging. 
The Briggs Robust=0.5 weighting was used to construct 
the continuum and molecular line images. 
Further analyses of data were performed using python 
\citep{2012ascl.soft08017R, 2013A&A...558A..33A}.

\section{Results and Analysis}
\label{sec:results}
\subsection{Dense structure identification}
To identify the compact structures in the 1.3~mm continuum 
image, we employed the 
{\tt astrodendro}\footnote{http://dendrograms.org/} 
algorithm to pre-select the compact components from the 
continuum image \citep{2013A&A...558A..33A}.  
There were compact structures detected at $>$5$\sigma$ 
significance, but were missed by {\tt astrodendro}.  We used 
the CASA-{\tt imfit} task to extract dense cores from  
the compact components identified from the {\tt astrodendro} 
analysis and also to recover the missed structures from the 
image.  
A detailed discussion about 
{\tt astrodendro} extraction method can be found in 
\cite{2008ApJ...679.1338R}. The input parameters of 
{\tt astrodendro} algorithm are described  below. The minimum 
pixel value, $\mathbf{min_{-}value}$, for which any pixel 
below this value is not considered, was set to 3$\sigma$, 
where $\sigma$ is the rms noise level of the continuum 
image. A value of 1$\sigma$ is chosen for 
$\mathbf{min_{-}delta}$, the minimum difference in the 
peak flux between neighboring compact structures for them 
to be considered as separate structures. We set the 
$\mathbf{min_{-}npix}$, which is the minimum number of 
pixels needed for structures to be considered as an 
independent entity, to the number of pixels within the 
FWHM of the synthesized beam. 
$\mathbf{min_{-}npix}$ ranges from 17 to 31 depending 
on the beam size of sources.  
Using the 1.3~mm continuum 
image, we have identified 4 to 10 cores in each of the 7 
clumps, and a total of 44 cores. The physical parameters 
of dense cores from CASA-{\tt imfit} are summarized 
in Table \ref{tbl:cores}, including coordinates, 
beam-convolved major axis ($\sigma_{maj}$), beam-convolved 
minor axis ($\sigma_{min}$), position angle (PA),  total 
integrated flux, beam-convolved ($R_{\rm eff}$) and 
beam-deconvolved effective radius ($R_{\rm eff}^{\rm dec}$).

\subsection{Mass determination}
\label{mass}
The gas mass ($M_{\rm gas}$) of dense cores are estimated 
using 1.3~mm continuum emission following the formula: 
\begin{equation}
\label{dust_mass}
M_{\rm gas}=\eta \frac{F_{\nu}\; d^{2}}{B_{\nu}(T_{\rm dust})\;\kappa_{\nu}},
\end{equation}
where $\eta$ ($\eta$ = 100) is the  gas-to-dust mass ratio, 
$d$ is the source distance, $F_{\nu}$ is the continuum 
integrated flux which has been corrected for primary 
attenuation, $B_{\nu}(T)$ is the Planck function at a dust 
temperature of $T_{\rm dust}$, and $\kappa_{\nu}$ is the dust 
opacity at a frequency of $\nu$. We adopt $\kappa_{\nu}$ = 0.84 
cm$^{2}$g$^{-1}$ by assuming 
$\kappa_{\nu}$ = 10(${\nu}$/1.2 THz)$^{\beta}$ cm$^{2}$g$^{-1}$ 
and $\beta$ = 1.5 \citep{1983QJRAS..24..267H}. 
The densities of clumps are higher than 10$^{4}$ cm$^{-3}$, which  
implies that gas and dust are likely well coupled toward 
these clumps \citep{2001ApJ...557..736G}.  We assumed 
$T_{\rm dust} = T_{\rm K}$, where $T_{\rm K}$ is 
kinetic temperature estimated from NH$_{3}$ data 
were retrieved from \cite{2012A&A...544A.146W} who 
performed observations of the NH$_{3}$ (J,K) = (1, 1) to 
(3, 3) inversion transitions  with the Effelsberg 100m 
Telescope. The $T_{\rm K}$ is in between 13 and 16 K, with   
typical uncertainty of about 10\%. For sources, 
AGAL024.314+00.086 and AGAL030.844+00.177,  
without NH$_{3}$ measurements, we adopt a temperature 
of $T_{\rm K} =$ 14 K.  
The derived gas masses of dense cores range from 1.4 to 
77.1 \Mo.  Massive cores ($\gtrsim$ 8 \Mo) are detected 
in all of the clumps, except for AGAL022.376+000.447 where 
the embedded dense cores have a maximum gas mass of 4.8 \Mo.

The column density ($N_{\rm H_{2}}$) of dense cores are 
computed with: 
 \begin{equation}
\label{column}
N_{\rm H_{2}}=\eta \frac{S_{\nu}^{\rm beam}}{B_{\nu}(T_{\rm dust})\;\kappa_{\nu} \; 
\Omega \; \mu \; m_{\rm H}},
\end{equation}
where $S_{\nu}^{\rm beam}$ is the peak flux density, 
$\Omega$ is the beam solid angle, $m_{\rm H}$ is the mass 
of an hydrogen atom, and $\mu$ ($\mu$=2.8) is the mean 
molecular weight of the interstellar medium 
\citep{2008A&A...487..993K}. 
The estimated $N_{\rm H_{2}}$ are between 
1.1$\times 10^{22}$ and 2.0$\times 10^{23}$ cm$^{-2}$, 
with a mean value of 5.7$\times 10^{22}$ cm$^{-2}$.

The uncertainty in  continuum flux is adopted to be a 
typical value of 10\% in the interferometer observations. 
The typical uncertainty in the kinematic distances is 
about 10\%, while in some cases it can be significantly 
larger due to the near-far kinematic distance ambiguity
\citep{2009ApJ...700..137R}. However, since all theses 
clumps appear as extinction images in the near-IR 
(i.e., they are infrared dark clouds), the near distance 
is likely the accurate distance.  
The $\eta$ is adopted to be 100 in this 
study, while its standard deviation is 23 (corresponding 
to a 1$\sigma$ uncertainty of 23\%) if we assume that it 
is uniformly distributed between 70 and 150 
\citep{1990ApJ...359...42D,2003A&A...408..581V,
2017ApJ...841...97S}. 
We adopted a conservative uncertainty of 28\% in 
$\kappa_{\nu}$ \citep[e.g.,][]{2017ApJ...841...97S}.  
Taking into account these uncertainties, we estimate an 
uncertainty of  44\%  and 32\% for gas mass and column 
density, respectively. However, one has to bear in mind 
that the uncertainties could be larger in  cases that 
distances are significantly higher.

\subsection{Dense cores and associated star formation}
Below, we summarize the general morphology of each of the 
7 clumps, and also comment on their CO outflow morphology. 
For each clump, dense cores are labeled in order of maximum
flux to minimum flux (Table \ref{tbl:cores}), e.g., MM1 to 
MM10, based the flux  prior to the primary beam correction.

\begin{figure*}[ht!]
\epsscale{1.1}
\plotone{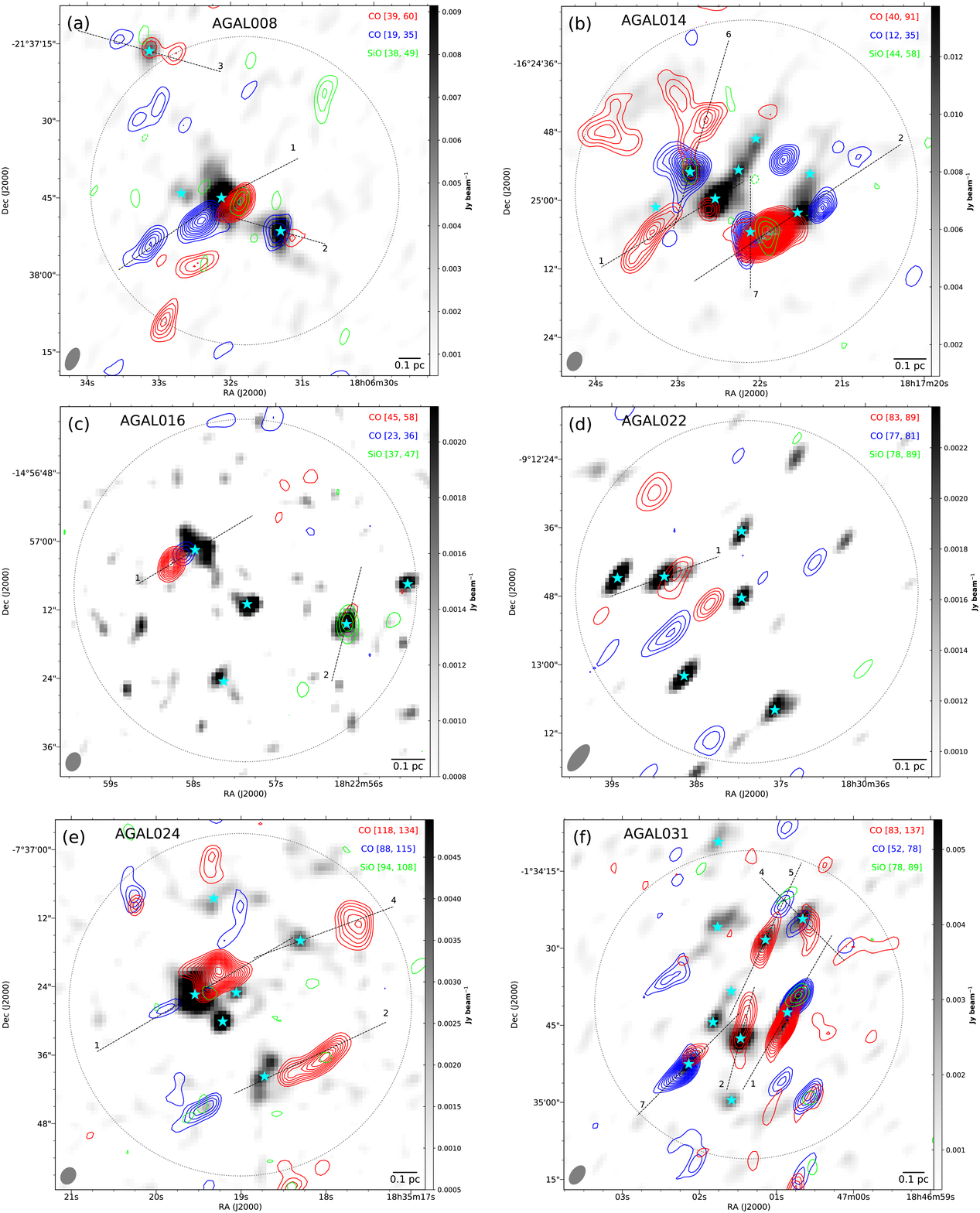}
\caption{
$\textbf{(a)}$ the redshifted (red contours) and 
blueshifted (blue contours) emission of CO outflows, 
and SiO velocity integrated intensity (green contours) overlaid 
on the 1.3~mm continuum emission. The blue and red contours are 
$\pm$(5, 7, 9 ...)$\times \, \sigma$, while the green  contours are 
$\pm$(3, 5, 7 ...)$\times \, \sigma$, where the $\sigma$ is 
the rms level. 
The dash black lines indicate the direction of 
outflows and the labelled corresponding to the dense cores 
numbers (e.g., 1 belongs to MM1). The integrated velocity 
range is shown in the upper right corner of image. The green 
stars represent the identified dense cores. The dash gray 
circle marks the FWHM of the SMA primary beam.  
The synthesized beam is shown in the bottom left corner 
of the image. $\textbf{(b)}$--$\textbf{(f)}$  same as
$\textbf{(a)}$, but for AGAL014, AGAL016,  AGAL022, 
AGAL024 and AGAL031, respectively. 
}
\label{fig:outflow}
\end{figure*}

AGAL008.691-00.401 (hereafter AGAL008) is a part of the 
clumpy structure that displays elongated emission running 
along the southwest-northeast direction as seen in the 
250 $\mu$m and 870 $\mu$m images (Figure \ref{fig:cont1}). 
AGAL008 breaks up into 4 dense cores in the SMA 1.3~mm 
continuum image, which are labeled as MM1 to MM4. The 
estimated gas masses of dense cores are between 8.1 and 
55.8 \Mo. All of dense cores are within the 20\% of  the 
primary beam of the SMA antenna, except for MM3 that is 
just outside the primary beam.  
CO outflows are detected toward MM1, MM2 and MM3. 
MM1 is associated with a high-velocity CO outflow with 
line-wing emission at velocities of 
$\geqslant$ 30 km s$^{-1}$ with respect to the source 
systemic velocity ($|v - v_{\rm lsr}| \geqslant$ 30 \kms)
(Figure \ref{fig:outflow}).

AGAL014.492+00.262 (hereafter AGAL014) shows an elongated 
feature oriented southeast-northwest (SE-NW) in APEX 870 
$\mu$m image, which breaks up into small clumpy 
structures in the SMA 1.3~mm continuum image 
(Figure \ref{fig:cont1}). Eight dense cores reside in  
the central  clumpy structure. Gas masses of dense 
cores are between 10.8 and 47.3 \Mo, four of which, 
MM1/2/6/7, are associated with CO outflows 
(Figure \ref{fig:outflow}). 
Both MM2 and MM6 are associated with high-velocity CO
outflows, which suggests that there are on-going star 
formation activities within these dense cores.

Figure \ref{fig:cont1} shows that the parent cloud of 
AGAL016.418-00.634 (hereafter AGAL016) appears as an 
elongated clumpy features oriented in the 
southwest-northeast direction. The 1.3~mm continuum 
image revealed five embedded dense cores within AGAL016, 
which has masses of 3.4 to 11.0 \Mo.   MM1 and MM2 are 
associated with outflows based on CO emission 
(Figure \ref{fig:outflow}).

For AGAL022.376+00.447 (hereafter AGAL022), there is an 
elongated emission running from east to west, as seen in 
the 250 $\mu$m and 870 $\mu$m images 
(Figure \ref{fig:cont1}). Using 1.3~mm continuum emission,
we identified 6 dense cores that contain masses between 
1.4 and 4.9 \Mo. The CO outflow signature is found 
toward MM1 (Figure \ref{fig:outflow}).

The 250 $\mu$m and 870 $\mu$m images reveal that the 
AGAL024.314+00.086 (hereafter AGAL024) is a part of a 
clumpy structure, which shows bright dust emission 
roughly running from southwest to northeast with a 
plateau around AGAL024. Using the 1.3~mm dust 
continuum image, we have 
identified 6 dense cores that contain gas masses between 
7.5 and 77.1 \Mo\ (Figure \ref{fig:cont1}).  There are 
three dense cores, MM1/2/4, appearing to have 
clear CO outflow emission (Figure \ref{fig:outflow}).

AGAL031.024+00.262 (hereafter AGAL031) is a part of 
circular clumpy structure that is clearly seen in the 
250 $\mu$m and 870 $\mu$m images 
(Figure \ref{fig:cont1}). AGAL031 breaks up into 10 
dense cores in the SMA 1.3~mm continuum image.  
Their masses range from 4.0 to 34.6 \Mo. 
CO outflows are detected in 5 out of 10 dense cores, 
including two cores, MM1 and MM5, with 
high-velocity CO outflows (Figure \ref{fig:outflow}).

The parent cloud of AGAL030.844+00.177 
(hereafter AGAL030) 
presents an elongated structure along the 
southeast-northwest direction (Figure \ref{fig:cont1}). 
AGAL030 breaks up into five dense cores  in the 1.3~mm 
image, with masses of 6.8 to 20.2 \Mo. There are no 
evidences of molecular outflow toward these dense cores
according to CO and SiO lines.

Overall,  44 dense cores have been identified using the 
1.3~mm continuum images. The identified dense cores have 
gas masses in the range of 1.4 to 77.1 \Mo\ with effective
radius between 0.02 and 0.11 pc. Of the 44 cores, 
29 cores have gas masses greater than 8 \Mo, 
while the other 15 cores are between 1.4 and 7.5 \Mo. 
The physical parameters of the identified dense cores 
are summarized in Table \ref{tbl:cores}.

\subsection{Dense cores structures}
\label{sec:density}
To study the physical structure of identified cores, we 
estimated core density distributions using the 1.3~mm 
continuum data. The detailed analysis of density 
structures of cores is described below. 

The SMA compact configuration gave projected baselines 
between $\sim$8 and $\sim$60 k$\lambda$ at the observing 
frequency ($\sim$230 GHz), which corresponds to angular 
physical size of 26\arcsec\ and  3.4\arcsec, respectively. 
If the dense core is internally heated, the dust 
temperature ($T_{\rm dust}$) and radius ($r$) can be expressed 
in the form of $T_{\rm dust} \propto r^{-a}$, where $a$ = 0.33 
\citep{1976ApJ...206..718S}. The observed flux density 
($F$) of the dust continuum emission along the line of 
slight can be written as 
$F \propto \int \rho \;T_{\rm dust} \;ds$ by assuming  
that the dust emission is optically thin and the density 
($\rho$) follows a power-law in radius 
$\rho \propto r^{-b}$. A power-law density profile has 
been proposed by theoretical studies 
\citep{1987ARA&A..25...23S,2003ApJ...585..850M}, and 
confirmed by observational investigations  
\citep{2000ApJ...537..283V,2001A&A...365..440M,
2005ApJ...633..535B}. 
For ($a + b$) $>$ 1, the observed flux density 
can be rewritten as $F \propto r^{-(a+b-1)}$. In the 
$u-v$ domain, following a Fourier transform, the observed 
flux density  can be rewritten as  
$A_{uv} \propto S_{uv}^{(a+b-3)}$ 
\citep{2000ApJ...529..477L, 2009ApJ...696..268Z}, where 
$A_{uv}$ is the visibility amplitude and 
$S_{uv}$ = $\sqrt{u^{2} + v^{2}}$ is the UV distance. 

We restrict our study to dense cores that have SNR 
higher than 9  and a roughly circular morphology in order 
to avoid contamination from unresolved nearby sources. 
The visibility data of dense cores is extracted from 
the map after removing the bright sources using clean 
components. We bin the extracted visibility data in step 
of 3.1 k$\lambda$ before we apply a non-linear 
least-squares minimization fitting 
\citep[LMFIT,][]{newville_matthew_2014_11813} to the 
visibility data in order to determine the density profile.

\begin{figure*}[ht!]
\epsscale{1.2}
\plotone{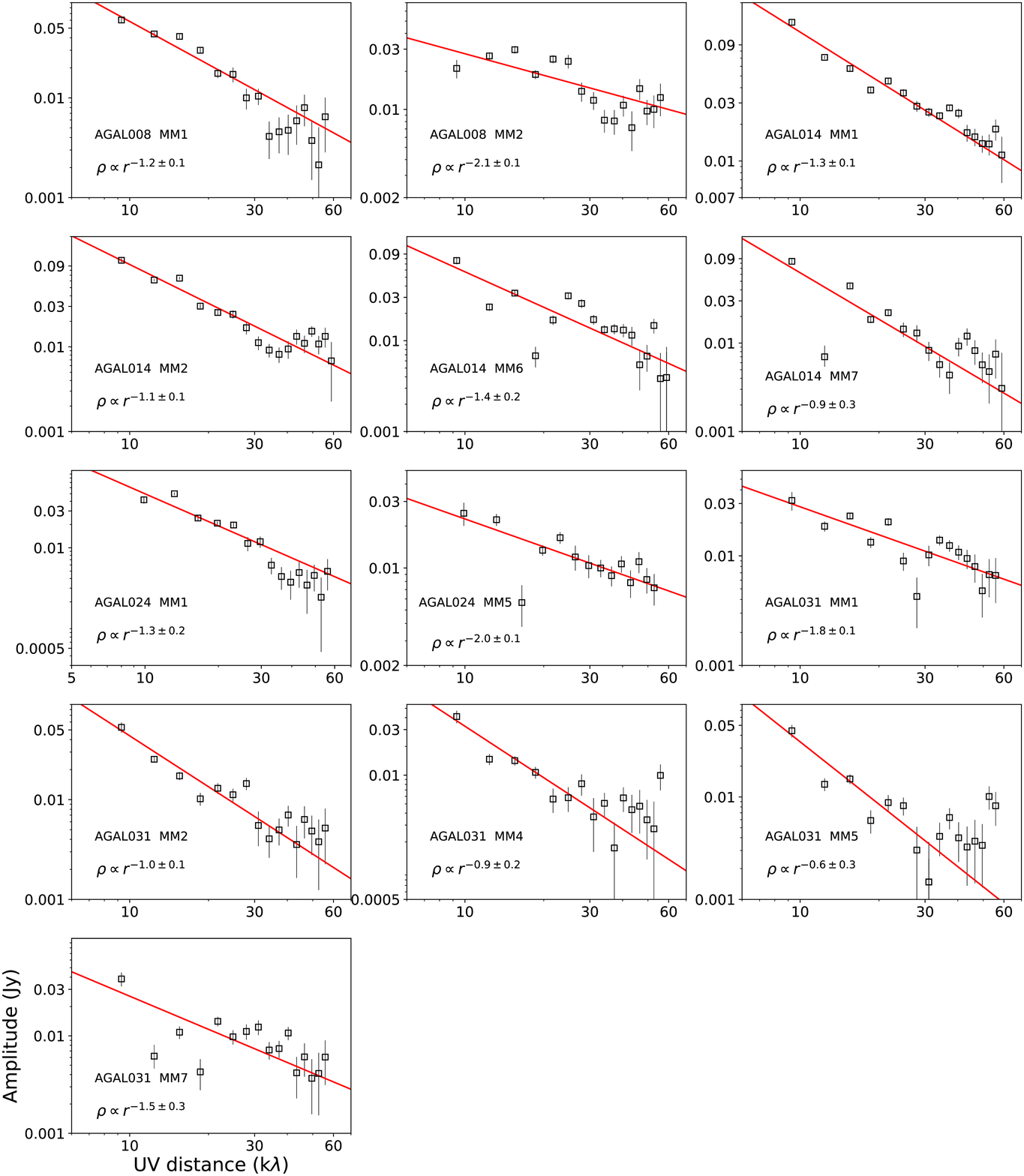}
\caption{Vector averaged amplitude distribution versus UV distance in the dense 
cores. The red solid line is the least-squared fitting result for the dense cores.}
\label{fig:uvamp}
\end{figure*}

\begin{figure*}[ht!]
\epsscale{0.7}
\plotone{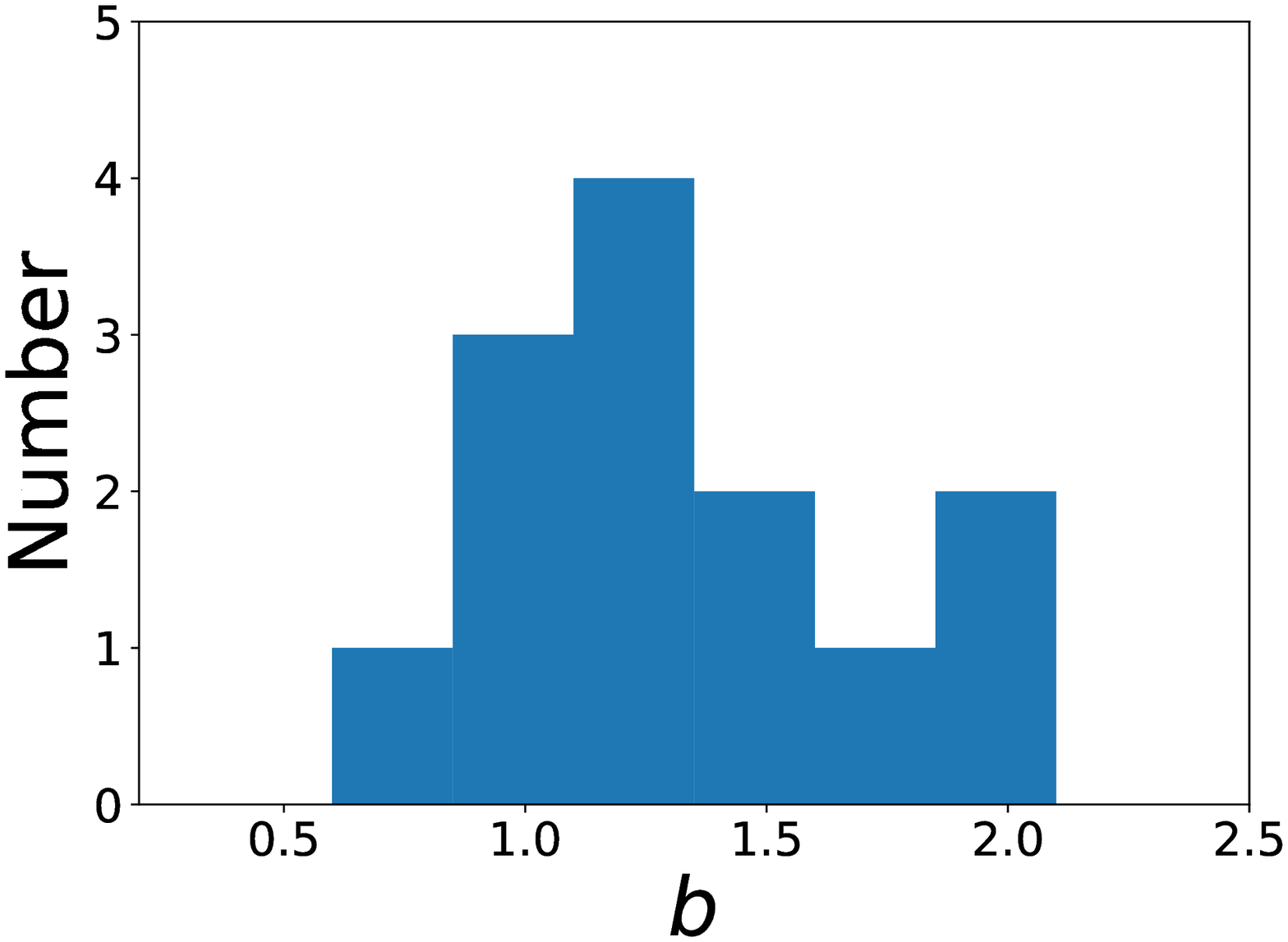}
\caption{The histogram is the power-law index of density distribution. 
}
\label{fig:dist_b}
\end{figure*}

The visibility amplitude versus 
the UV distance with the fitting results are shown in 
Figure  \ref{fig:uvamp}. The amplitude is vector averaged 
over concentric annulus around the source in the $u-v$ 
plane. The large scatter  at long UV distances  are 
mostly  due to the limited signal-to-noise, while it can 
not be ruled out  the possibility that the cores host 
multiple unresolved compact structures. In some cases, 
the fitting is not as robust due to large scatters, 
especially for 3 cores, i.e., AGAL008-MM2, 
AGAL014-MM7 and AGAL031-MM7. 
Assuming $a$ = 0.33, the best fitted density power-law 
indexes ($b$) of dense cores are between 0.6 and 2.1, 
with a mean value of 1.3. Figure \ref{fig:dist_b} shows 
the density power-law indexes distribution for dense cores.

In previous studies, \cite{2000ApJ...537..283V}  found 
that the power-law density distribution has a slope of 
$b$ = 1.0 $-$ 1.5 through measured of a sample of 14 
ultracompact HII regions and hot cores on 
10$^{2}$ $-$ 10$^{5}$ AU scales. 
On the other hand, \cite{2002ApJS..143..469M} 
via observations of 51 massive dense cores 
($\sim$3.3$\times$10$^{4}$ AU scale) 
found  that b ranges from 0.75 to 2.5 with a mean value 
of $\langle b \rangle$ = 1.8 $\pm$ 0.4. This  is similar 
to the value of $\langle b \rangle$ = 1.6  reported in 
\cite{2005ApJ...633..535B}, who studied 69 massive cores 
on scale of $\sim$10$^{4}$ AU.  In a more recent study,  
\citep{2012ApJ...754....5B} found that for a sample of 42 
IRDC cores ($\sim$2$\times$10$^{4}$ AU scale), b was 
about 1.6.  As a comparison, low-mass protostellar cores 
(10$^{3}$-10$^{5}$ AU scale) have  $b$ = 1.5 $-$ 2.0 
\citep{2001A&A...365..440M}, 
$b$ = 1.5 $-$ 2.5 \citep{2015ApJ...798..128T}  
and $\langle b \rangle$ = 
1.6 $\pm$ 0.3 \citep{2002ApJ...575..337S}. 
Our results are in agreement with these investigations, 
as we find no significant differences between power-law 
indexes of our 70 $\mu$m dark dense cores and the 
above mentioned high- and low-mass objects.

\subsection{Outflow properties} \label{Outflowparameters}
Outflow signatures, as revealed by CO emission, are 
detected toward 17 identified dense cores.  The CO 
outflow emission is detected in all of the clumps, except 
for AGAL030. The maximum detected velocities 
($|v - v_{\rm lsr}|$) for the CO 
outflow emission range from 5 up to 56 km s$^{-1}$ with 
respect to the source systemic velocity. There are five 
dense cores associated with high velocity CO outflow 
emission ($|v - v_{\rm lsr}| \geqslant$ 30 \kms) within a 
small spatial scale of $\leqslant$ 0.22 pc, while the 
rest of the dense cores are associated with low-velocity 
CO outflows ($|v - v_{\rm lsr}| <$ 30 \kms).  The 
velocity-integrated intensities of CO emission are shown 
in Figure \ref{fig:outflow}, while the blue and red color 
contours represent the red-shifted and blue-shifted 
emission components, respectively. From the maps, we note 
that some of the cores show well collimated CO outflows 
with a bipolar morphology, while others  are dominated by 
either the blue- or red-shifted lobes. The SiO emission has 
been detected toward 5 massive and 1 intermediate-mass 
dense cores that are associated with CO outflows 
(Figure \ref{fig:outflow}).  The SiO emission 
is much fainter and with a narrower velocity range 
than that of the CO emission. 

To study outflow properties, we estimated the physical 
parameters for each outflow, including mass, momentum, 
energy, dynamical timescale, outflow rate, as well as 
accretion rate. Assuming local thermodynamic equilibrium, 
we first estimate the CO column density following 
\cite{2015PASP..127..266M}: 
\begin{multline}
\label{outflow-mass}
N_{\rm CO}({\rm cm^{-2}}) =  1.08 \times 10^{13} (T_{\rm ex} + 0.92)\; 
exp\left( \frac{16.59}{T_{\rm ex}} \right)\\
\int \frac{\tau_{12}}{1-e^{\tau_{12}}} T_{\rm B}dv,
\end{multline}
where $N_{\rm CO}$ is the CO column density, $dv$ is the 
velocity interval in km~s$^{-1}$, $T_{\rm ex}$ is the 
line excitation temperature, $\tau_{12}$ is the optical 
depth, and $T_{\rm B}$ is brightness temperature in K. 
The outflow mass ($M_{\rm out}$), momentum ($P_{\rm out}$), 
energy ($E_{\rm out}$), dynamical age ($t_{\rm dyn}$), and outflow 
rate ($\dot{M}_{\rm out}$) can be estimated with  
\citep{1983ApJ...265..824B,1992A&A...261..274C} 
\begin{equation}
\label{Mout}
M_{\rm out} = {d^{2}} \left[\frac{\rm H_{2}}{\rm CO}\right] 
\overline{m}_{\rm H_{2}} \int_{\Omega} N_{\rm CO}(\Omega) d\Omega,
\end{equation}
\begin{equation}
\label{Pout}
P_{\rm out} = M_{\rm r} v_{r} + M_{b} v_{\rm b},
\end{equation}
\begin{equation}
\label{Eout}
E_{\rm out} = \frac{1}{2} M_{\rm r} v_{\rm r}^{2} + \frac{1}{2} M_{\rm b} v_{\rm b}^{2},
\end{equation}
\begin{equation}
\label{tdyn}
t_{\rm dyn} = \frac{l_{\rm out}}{(v_{\rm max(b)} + v_{\rm max(r)})/2},
\end{equation}
\begin{equation}
\label{Moutrate}
\dot{M}_{\rm out} = \frac{M_{\rm out}}{t_{\rm dyn}},
\end{equation}
where $\Omega$ is the total solid angle that the flow 
subtends, $d$ is the source distance, $l_{\rm out}$ is the CO 
outflow physical length, $v_{\rm max(b)}$ and $v_{\rm max(r)}$ 
are  the maximum velocities of CO blue-shifted and 
red-shifted emission, respectively. $M_{\rm r}$ and $M_{\rm b}$ 
are the gas masse of CO outflows at blue-shifted ($v_{\rm b}$) 
and red-shifted ($v_{\rm r}$) velocities, respectively.   
Here, we adopt the CO-to-H$_{2}$ abundance of 10$^{-4}$, 
$\left[ \frac{\rm H_{\rm 2}}{\rm C\rm O}\right]$ = 10$^{4}$ 
\citep{1987ApJ...315..621B}, mean mass per hydrogen atom 
$\overline{m}_{\rm H_{2}}$ = 2.33, and assume that  CO 
emission is optically thin in the line wing and that the excitation 
temperature of outflow gas equals to the temperature 
($T_{\rm K}$) estimated from the NH$_{3}$ emission 
\citep{2012A&A...544A.146W}.  As mentioned in Section 
\ref{mass}, a temperature of $T_{\rm K} = $14 K was assumed 
for the sources without NH$_{3}$ observations (AGAL024 and 
AGAL030). The primary beam correction have been applied 
to the CO velocity integrated intensity for estimating 
the outflow masses, while the  inclination of the outflow 
axis with respect to the line of sight have not been 
corrected in the derivation of the outflow parameters 
(see Section \ref{inclination} for the discussion of inclination).

The derived outflow masses range from 0.008 to 0.17 
M$_{\odot}$ with a mean value of 0.06 M$_{\odot}$. 
The outflow momenta are between 0.04 and 2.7 
M$_{\odot}$ km s$^{-1}$, while the outflow energies are 
in the range of 0.1 to 34.8~M$_{\odot}$ km$^{2}$ s$^{-2}$. 
Using the outflow mass and dynamical time, we find the 
outflow ejection rates are between 0.6~$\times$~10$^{-6}$ 
and  2.1~$\times$~10$^{-5}$ M$_{\odot}$ yr$^{-1}$, with a 
mean value of 0.5~$\times$~10$^{-5}$ M$_{\odot}$ yr$^{-1}$.
In general, the most massive cores tend to have high 
outflow mass, momentum, energy, outflow ejection rate and 
accretion rate within the same clump. The ratios of 
$M_{\rm out}$/$M_{\rm core}$  are around 0.003, which is  
lower than the value of 0.04 reported by 
\cite{2002A&A...383..892B}, who studied a sample  of much 
massive (10$^{2}$ -- 10$^{5}$ $M_{\odot}$) and evolved 
objects with relatively low spatial resolution 
($\sim$11\arcsec). 
It should be noted that the outflow mass estimated here is 
only a lower limited since the CO emission suffers from 
missing flux and is likely to be optically thick 
\citep[see, e.g.,][]{2014ApJ...783...29D}, as well as outflow 
parameters. The derived outflow parameters are 
summarized in Table \ref{tbl:outflow}.

Assuming that the outflows are powered by winds that 
driven by accretion disks 
\citep[e.g.,][]{2003ApJ...599.1196K,2003ApJ...585..850M} 
and that there is efficient mixing at the wind/molecular gas 
interface \citep{2000prpl.conf..867R}, the conservation 
of momentum can be used to approximate the 
relation between outflows and winds.  In this case, the 
mass-loss rate of the wind can be estimated by 
$\dot{M}_{\rm w} v_{\rm w} = P_{\rm out} / t_{\rm dyn}$, 
where $\dot{M}_{\rm w}$ and $v_{\rm w}$ are 
mass-loss rate and the 
velocity of the wind, respectively. We assume a wind 
velocity of 500 km s$^{-1}$ and a ratio of 3 for the 
mass-loss rate of wind to accretion rate 
\citep{1998ApJ...502L.163T,2000prpl.conf..789S}. 
The derived accretion rates are between 
1.8~$\times$~10$^{-8}$ and 
2.3~$\times$~10$^{-6}$ \Mo\ yr$^{-1}$, with 
a mean value of 4.4~$\times$~10$^{-7}$ \Mo\ yr$^{-1}$. 
The derived accretion rates are similar to that of 
24~\um\ dark sources \citep{2015ApJ...805..171L}, while 
they are two or three order magnitude lower than that of 
more evolved sources; the typical value is several 
10$^{-4}$ \Mo\ yr$^{-1}$ for HMPOs 
\citep{2002A&A...383..892B,2005ApJ...625..864Z,
2009ApJ...696...66Q} and $\sim$10$^{-5}$ \Mo\ yr$^{-1}$ 
for 24~\um\ bright sources 
\citep{2011ApJ...735...64W,2015ApJ...804..141Z}.

\begin{figure*}[ht!]
\epsscale{1.2}
\plotone{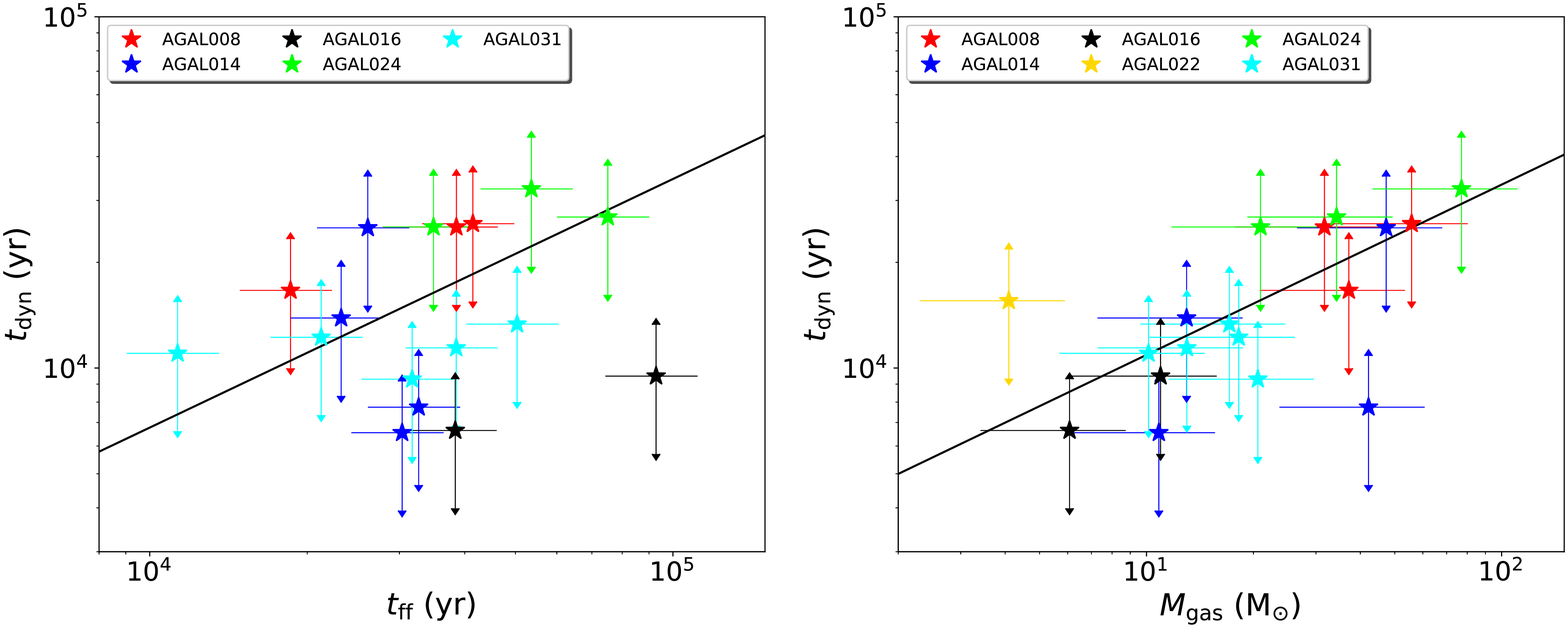}
\caption{Left: the free-fall time versus outflow dynamical time. 
The Spearman-rank correlation test returns a correlation coefficient 
of 0.59, while the best Least-squares fitting result is $y = 10x^{0.7}$ 
(black solid line). 
Right: the dense core mass versus outflow 
dynamical time. The correlation coefficient is about 0.62 from the 
Spearman-rank correlation test. The black solid line is the 
best least-squares fitting result $y = 3570x^{0.5}$. 
The errorbars values of t$_{dyn}$ correspond to the mean inclination 
angle ($\langle \theta \rangle \approx 57.3^{\circ}$) correction. 
The errorbars of $t_{\rm ff}$ and $M_{\rm gas}$ correspond to their 
uncertainties.
}
\label{fig:outflow-para}
\end{figure*}

To compare the free-fall timescale and the outflow 
dynamical timescale, we have computed the average 
volume density in dense cores, 
$n_{\rm H_{2}} = 3M_{\rm gas}/(4\pi R^{3})$, and 
the free-fall time, 
$t_{\rm ff} = \sqrt{3 \pi/(32\,G\,n_{\rm H_{2}}) }$. 
Here, $n_{\rm H_{2}}$ is dense core averaged-volume density, 
$R$ is the effective radius, $t_{\rm ff}$ is the free-fall timescale 
and $G$ is the gravitational constant. 
The outflow dynamical timescales show a correlation 
with the free-fall timescales 
(Figure \ref{fig:outflow-para}). The Spearman-rank 
correlation test, which assesses monotonic relationships, 
returns a coefficient of 0.59. This indicates a moderate 
correlation between the two timescales. They are in same 
order of magnitude ($\sim$ few 10$^{4}$ yr) and is 
relatively smaller than the typical outflow dynamical  
timescale of $\sim$10$^{5}$ yr in more evolved objects, 
such as HMPOs 
\citep{2002A&A...383..892B,2005ApJ...625..864Z}. 
The shorter timescale is  consistent with the fact that 
our sample  is at a very early evolutionary stage of 
star formation.

We find a positive correlation between outflow dynamical 
timescale and  gas mass of dense cores, with a correlation
coefficient of 0.62 estimated from Spearman-rank correlation
test. The massive cores tend to have a longer outflow 
dynamical timescale than that of less massive cores. This 
trend suggests that the more massive cores had longer 
accretion history that allows it to assemble more mass. 
This also implies that the most massive cores may form 
earlier than the less massive cores.  A confirmation of 
this trend from a more sensitive and larger sample is 
still needed.

\section{Discussion}

\subsection{Dense cores dynamical state}
\label{virial}
We computed  the  virial mass for both clumps and dense 
cores  following:
\begin{equation}
M_{\rm vir}=3k \frac{\sigma_{v}^{2} R}{G} ,
\end{equation}
where $M_{\rm vir}$ is the virial mass, $\sigma_{v}$ is 
the velocity dispersion, $R$ is the effective radius, 
$G$ is the gravitational constant, and the correction 
factor $k$ = (5-2$b$)/(3-$b$) depends on the density 
profile $\rho \propto r^{-b}$ \citep{1988ApJ...333..821M}. 
Here, we use the derived density profile $b$ for dense cores, 
and a mean value of $\langle b \rangle$ = 1.3 for dense 
cores without $b$ measurements (see 
Section \ref{sec:density}). 
The virial parameter is computed using 
$\alpha = M_{\rm vir}/M_{\rm gas}$, where $M_{\rm gas}$ 
is the gas mass.

We used the C$^{18}$O line width to estimate the velocity 
dispersion for dense cores, as was done for condensations 
in IRDC G28.34+0.06 \citep{2015ApJ...804..141Z}. This 
line was chosen over other lines observed in the SMA 
SWARM bandwidth because it is bright and less confused by 
outflow motions and/or severe depletion than H$_{2}$CO, 
CH$_{3}$OH, HCN and HCO$^{+}$. 
We extracted the average spectrum 
of C$^{18}$O within dense cores and fitted the Gaussian 
profile to measure the observed line width (full width  
half maximum, FWHM), $\bigtriangleup v_{\rm obs}$. 
The derived C$^{18}$O line widths are systemically 
narrower than the $^{13}$CO line, with a mean 
$\bigtriangleup v_{\rm obs}$ of 1.7 km s$^{-1}$ and 
2 km s$^{-1}$ for C$^{18}$O and $^{13}$CO, respectively.  
This could be due to the fact that the optical depth of 
$^{13}$CO is larger than that of C$^{18}$O 
\citep[e.g.,][]{2016A&A...591A.104H}, or the $^{13}$CO 
line widths are  affected by outflows 
\citep{2018ApJS..237...22S}.
Therefore, the line width from C$^{18}$O is better suited 
for estimating of the virial parameters of the identified 
dense cores.

The velocity dispersion $\sigma_{v}$ contains both 
non-thermal and thermal motions from particles of mean 
mass contributions.  To estimate the velocity dispersion, 
we first subtract the C$^{18}$O thermal motions from the 
observed line width using 
$\sigma_{\rm nt}^{2}$ = $\sqrt{\sigma_{\rm obs}^{2} - \sigma_{\rm th}^{2}}$, 
where $\sigma_{\rm obs}$ is the channel-deconvolved 
line width, $\sigma_{\rm obs}$  = $\left(\bigtriangleup v_{\rm obs}^{2} - 
\bigtriangleup v_{\rm ch}^{2} \right)^{0.5}  \left( 2\sqrt{2ln(2)} \right)^{-0.5}$,  
and $\bigtriangleup v_{\rm ch}$ is the spectral 
resolution. The molecular thermal motion 
$\sigma_{\rm th}$ = 9.08 $\times$ 10$^{-2}$ km s$^{-1} 
\left(\frac{T}{\rm K}\right)^{0.5}  
\left(\frac{m}{m_{\rm H}}\right)^{-0.5}$ 
\citep{1983ApJ...270..105M},
where $m$ is the molecule weight, $m_{\rm H}$ is the 
hydrogen mass, and $T = T_{\rm K}$ is the gas temperature. 
We used this 
equation to calculate the thermal motion (sound speed 
$c_{s}$) of the free particle of mean mass assuming a mean 
mass of gas of 2.37\,$m_{\rm H}$ 
\citep{2008A&A...487..993K}. Finally we estimated the 
velocity dispersion using 
$\sigma_{v}$ = $\sqrt{\sigma_{\rm nt}^{2} + c_{s}^{2}}$.

\begin{figure*}[ht!]
\epsscale{1.2}
\plotone{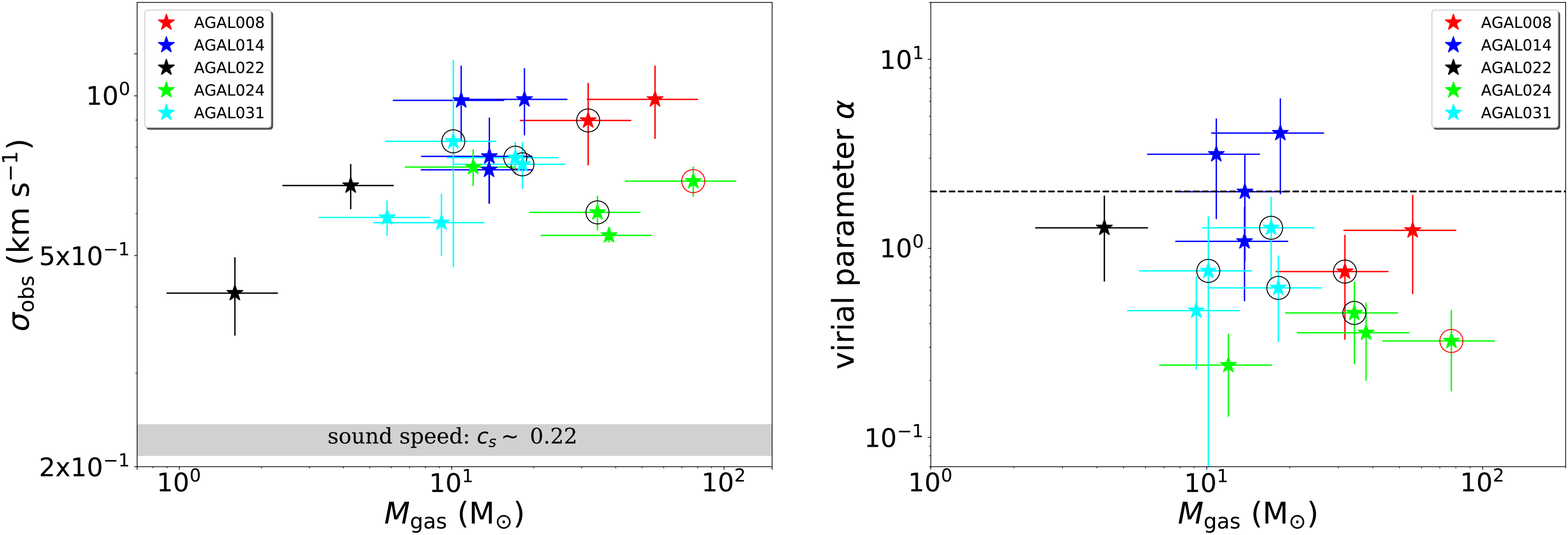}
\caption{Left: the C$^{18}$O line width, after 
deconvolution with the channel width, versus gas mass of 
dense cores. The grey shadow is the sound speed.  
Right: the virial parameter, $\alpha$, against 
gas mass of dense cores. 
The red and black circles represent the cores associated 
with high-  (velocity $\geqslant$ 30 km s$^{-1}$ with respect to 
the source systemic velocity) and low-velocity 
(velocity $<$ 30 km s$^{-1}$) CO outflow, respectively. 
}
\label{fig:virial}
\end{figure*}

We restrict the virial analysis to 15 dense cores 
whose C$^{18}$O spectra have fitted SNR higher than 5.  
Figure~\ref{fig:virial} shows the $\sigma_{\rm obs}$ 
versus gas mass for the dense cores, while the red and 
black circles represent the high-  and low-velocity  CO 
outflows, respectively.   Although number of data points 
is limited, we find no significant relation between 
$\sigma_{\rm obs}$ and dense cores with and without 
associated CO outflows.  Therefore, the line width 
derived by C$^{18}$O is a reliable parameter for 
estimating the virial parameter, $\alpha$. 
The derived virial parameters of dense cores range from 
0.2 to 4.1 with a mean value of 1.1. 
After propagating uncertainties from gas mass and virial 
mass, we find the uncertainties for the virial parameters 
are $\sim$54\%.  

Non-magnetized cores with $\alpha <$ 2, $\alpha \sim$ 1 
and $\alpha <$ 1 are considered to be gravitationally 
bound, in hydrostatic equilibrium and gravitationally 
unstable, respectively. Considering the uncertainty of 
virial parameters, we find that the virial parameters of 
8 cores are smaller than 1, of 5 cores are between 1 and 
2, and of 2 cores range from 2 to 4.1.  This  indicates 
that most of the dense cores are gravitationally bound. 
The virial parameters are not significantly different 
for  dense cores associated with and without CO outflows. 
Figure \ref{fig:virial} shows the virial parameters 
versus the gas mass of dense cores. We noted that the 
most massive cores tend to have lower virial parameters. 
Previous investigations of parsec scale massive clumps 
and low-mass star formation regions have found similar 
trend between virial parameter and mass 
\citep{2008ApJ...672..410L,2009ApJ...696..298F,
2015MNRAS.452.4029U}. Several recent studies find that the 
observation anti-correlation between the virial parameter 
and the mass is subject to a few observational biases 
\citep{2013ApJ...779..185K,2018MNRAS.479.2112B,
2018A&A...619L...7T}.

To study the virial state of the dense clumps, we use the
NH$_{3}$ (1, 1) line widths \citep{2012A&A...544A.146W}.  
Both AGAL024 and AGAL030 are not been considered in the 
analysis of clump virial state since they are no 
observations of NH$_{3}$ nor  N$_{2}$H$^{+}$ in the low-J 
transitions. The NH$_{3}$ (1, 1)  velocity dispersion  
range from 0.6 to 1.2~km~s$^{-1}$ for these clumps. The 
derived virial parameters of the clumps are between 0.44 
and 0.61, which indicates that these clumps are 
gravitationally unstable.

Our analysis does not include magnetic fields and 
external pressure. The magnetic fields supply additional 
support to counteract gravity 
\citep{2019FrASS...6....3H}.
Since these cores are embedded in dense clumps, the 
external pressure with the cloud provides additional 
force against the internal support, which could decreases 
the virial parameters.  Therefore, these effects may play 
an important role in the balance between the external 
pressure, gravitational potential energy and internal 
kinetic energy \citep{2015MNRAS.450.1094P,
2015ApJ...804..141Z,2017ApJ...846..144K}.

\subsection{Fragmentation}
Fragmentation exists on various spatial scales from giant 
molecular clouds down to dense cores. In this study we 
investigate how the pc-scale massive 70~$\mu$m dark 
molecular clumps fragment into 0.1 pc-scale dense cores 
that might form massive stars.

With the high spatial resolution and mass sensitivity, 
we can directly compare the detected dense cores with 
predictions of  Jeans fragmentation 
\citep{2012sse..book.....K}.  
The APEX 870~$\mu$m continuum and NH$_{3}$ (1, 1) 
emission are used to compute the 
Jean length and Jeans mass under pure thermal support, 
and both thermal and non-thermal  support 
(turbulent support). 
\begin{multline}
\label{Jeanslen}
\lambda_{\rm J}= \sigma \left(\frac{\pi}{G \rho} \right)^{1/2} \\
=  0.06\, \rm pc \left[\frac{\sigma}{0.188\; \rm km\; \rm s^{-1}} \right] 
\left[\frac{n_{\rm H_{2}}}{10^{5} \; \rm cm^{-3}}\right]^{-1/2},
\end{multline}
Assuming a spherical symmetry with a radius of 
$\lambda_{\rm J}$/2 the Jeans mass can be estimated by: 
\begin{equation}
\label{Jeansmass}
M_{\rm J}= \rho \lambda_{\rm J}^{3} 
= 0.8\, \rm M_{\odot} \left[\frac{\sigma}{0.188\; \rm km\; \rm s^{-1}} 
\right]^{3} \left[\frac{n_{\rm H_{2}}}{10^{5}\; \rm cm^{-3}} \right]^{-1/2},
\end{equation}
Where $\lambda_{\rm J}$ is the Jeans length,  
$M_{\rm J}$ is the Jean mass, $n_{\rm H_{2}}$ is the 
dense core averaged-volume density, and 
$\sigma$ is the  velocity dispersion. If the dense cores 
are supported by pure thermal motions, the $\sigma$ 
equals to the sound speed. If the dense cores are 
supported by turbulent motions, then 
$\sigma$ equal to 
$\sigma_{v}$ = $\sqrt{\sigma_{\rm nt}^{2} + c_{s}^{2}}$. 
We also calculated the Jeans number $N_{\rm J}$, which 
is given as 
\begin{equation}
\label{Jeansunmber}
N_{\rm J} = \frac{M_{\rm clump}}{M_{\rm J}}
\end{equation}
Where $M_{\rm clump}$ is the total mass of compact clumps 
toward the region of our observations.  
The larger Jeans number means the more possibility of 
fragmentation. 
The estimated Jeans mass, Jeans length, and Jeans 
number are summarized in Table \ref{tbl:Jean}. 
The gas masses of dense cores are estimated from the 1.3~mm 
dust emission (see section \ref{mass}). The separations 
of dense cores are the projected distance on the sky of 
the nearest neighbor cores.

\begin{figure*}[ht!]
\epsscale{1.2}
\plotone{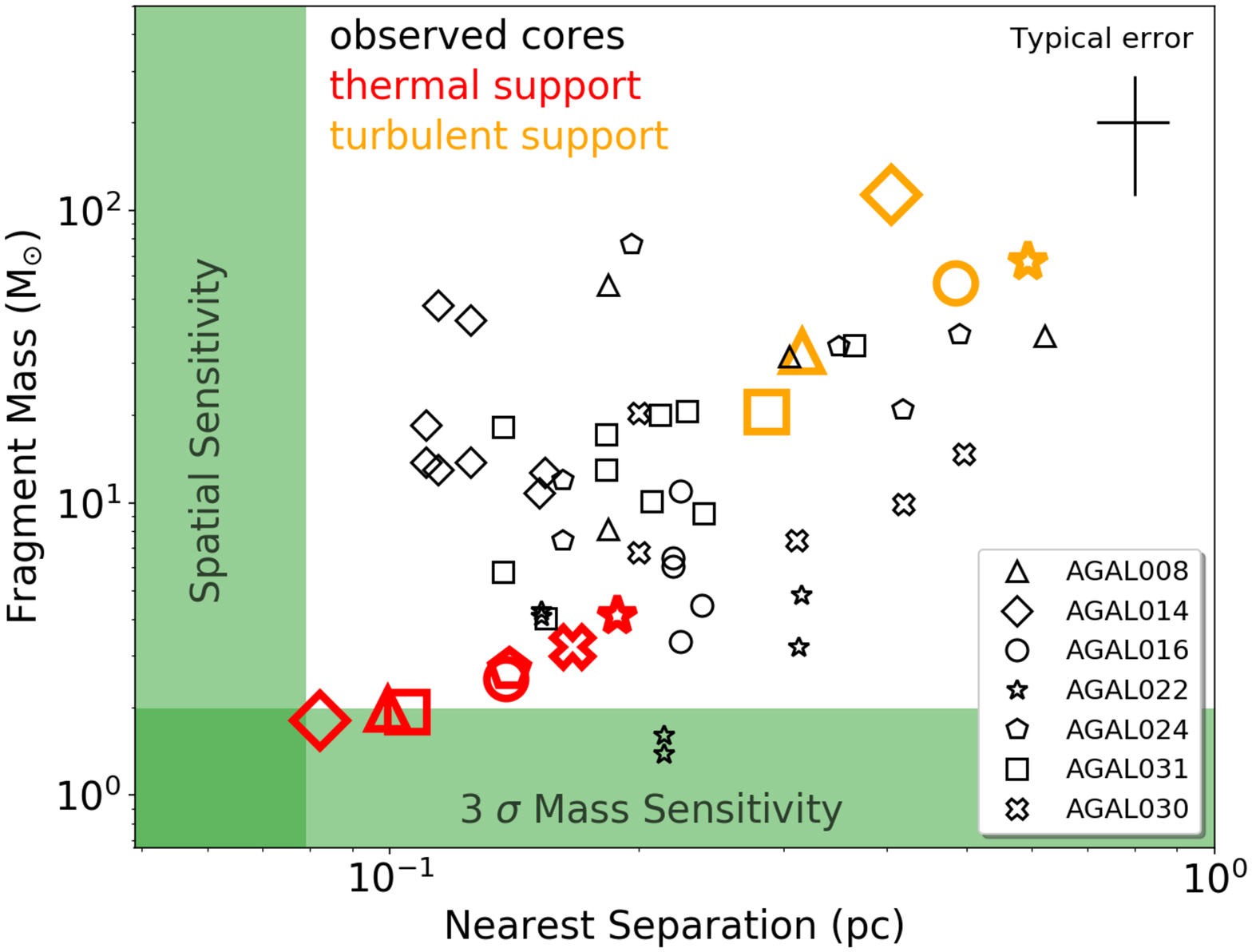}
\caption{Fragment mass versus  projected separation on the sky 
of the nearest neighbor fragments. 
The black symbols are the fragments identified by SMA observations. 
The black cross is the typical uncertainly of the observed fragments. 
The red symbols represent the prediction of thermal Jeans fragmentation. 
The orange symbols represent the prediction of turbulent Jeans 
fragmentation, 
where the turbulent component is from the clump non-thermal 
NH$_{3}$ line width. 
The green shaded regions show the mass sensitivity and spatial 
resolution limit of the observations. 
}
\label{fig:Jeansfrag}
\end{figure*}

In the case of pure thermal fragmentation 
(Figure \ref{fig:Jeansfrag}),  the Jeans masses are 
much smaller than the observed massive 
cores and more consistent with the low-mass cores.   
In AGAL022 that only detected intermediate- and 
low-mass cores, the Jeans mass and length are roughly 
consistent with the observed values. 
This is in agreement with the studies of 
\cite{2013ApJ...762..120P,2015MNRAS.453.3785P} 
and \cite{2018ApJ...853....5P}, who investigated a sample 
of intermediate- and low-mass cores.  The pure thermal 
Jeans fragmentation fails to explain the massive cores that 
contain masses much greater than the thermal Jeans mass.  
The Jean numbers of pure thermal fragmentation are 
significantly larger than the observed numbers of dense cores, 
and leads to the ratios of Jean number to observed number 
of 0.01-0.03.  These ratios are similar to what 
\cite{2018ApJ...853....5P} found in Perseus (0.06), which 
suggests that pure thermal motions do not support large clumps. 
Additional support, such as turbulence and/or magnetic fields, 
may be needed to counteract gravitational
collapse and further fragmentation.  On the other hand, 
we cannot rule out that these massive cores have a higher 
multiplicity that is not resolved due to limited spatial resolution 
\citep[e.g.,][]{2015ApJ...804..141Z,2018ApJ...855...24P}.

We also study the Jeans fragmentation with regards to the 
non-thermal motion together with the thermal support 
(also known as turbulent Jeans fragmentation). Since 
both AGAL024 and AGAL030 have no reliable line width 
information, they are not considered in this analysis. 
The predicted turbulent Jean masses are approximately 
consistent with the observed massive cores (see 
Figure \ref{fig:Jeansfrag}). This indicates that the turbulent 
support may be needed for these massive cores, which  is 
consistent with the results of investigations of massive dense 
cores
\citep[e.g.,][]{2009ApJ...696..268Z,2011A&A...530A.118P,
2011ApJ...733...26Z,2014MNRAS.439.3275W,
2015ApJ...805..171L,2015ApJ...804..141Z}.
The Jean number of turbulent Jeans fragmentation are slightly 
larger than or similar to the observed numbers of dense cores, 
which implies that the turbulent motions might play a role in 
supporting larger clumps.

The velocity dispersion in the dense cores 
($R_{\rm eff}^{\rm dec} \sim$ 0.05 pc) is smaller than their 
natal clumps ($R_{\rm eff}^{\rm dec} \sim$ 0.65 pc).  The 
former is dominated by transonic motions (i.e., as 
derived by C$^{18}$O emission), while the latter is 
dominated by supersonic motions (as derived by NH$_{3}$). 
Such variations in the line width from the clump scale down 
to the dense core scale toward massive star formation regions 
have been reported in previous studies 
\citep{2008ApJ...672L..33W,2015ApJ...804..141Z}. 
The observed supersonic line widths at clump scales could 
be due to  the effect of a poor linear resolution; 
spatially unresolved motions within the telescope beam 
broaden the observed line widths. The line widths 
decrease  from clump to dense core scales also could be 
due to the dissipation of turbulences.  Since C$^{18}$O 
(a critical density $c_{\rm cr} \sim $ 10$^{4}$ cm$^{-3}$) 
is more easily depleted via freezing onto dust grains than that 
of NH$_{3}$ 1-1 ($c_{\rm cr} \sim $ 10$^{3}$ cm$^{-3}$),  
it is possible that NH$_{3}$ and C$^{18}$O are tracing 
different gas components.

The gas masses of the majority (90\%) of the identified dense 
cores are greater than 4 \Mo, which is higher than the gas 
mass in the predictions of competitive accretion model 
\citep{2002MNRAS.336..659B,2006MNRAS.370..488B}. 
We did not find a population of low-mass dense cores within 
these massive clumps, which certainly could be due to the lack 
of mass sensitivity and the spatial resolution of the observations.  
In addition, the dense cores are not in a virial equilibrium 
(see Figure \ref{fig:virial}), which is not consistent with the 
turbulent core model which hypothesizes that the dense cores are 
in a virial equilibrium due to  turbulent support that  counteracts 
gravity \citep{2002Natur.416...59M,2005Natur.438..332K,
2007ApJ...656..959K}.  However, we caution that the 
identified massive cores might fragment into less massive 
cores in higher spatial resolution observations 
\citep[e.g.,][]{2015ApJ...804..141Z}. In addition, the presence 
of strong magnetic fields provides additional internal support 
and helps to virialize the core.

To study the relationship of the number of cores fragmented from 
the clump with the clumps properties, we have computed the
clump parameters using APEX 870~\um\ emission, including 
column and volume densities, Mach number, gas mass, virial 
parameter and density distribution. Comparisons of fragments 
with the clump properties show no apparent correlation. 
This could be due to either the small sample size, or the 
fact that the fragmentation is dominated by a combination of 
multiple physical processes (e.g., turbulence, magnetic fields, 
protostar feedback, and density distributions).

\subsection{Inclination corrections}\label{inclination}
Since we can not derive the inclination of the outflow axes, 
the parameters in Table \ref{tbl:outflow} have not been corrected 
for their inclination. 
Assuming a outflow inclination angles $\theta$ ($\theta$ = 0$^{\circ}$ 
correspond to outflow along the line of sight, $\theta$ = 90$^{\circ}$  
correspond to the outflow perpendicular the line of sight), we can 
estimate the correction factors for the outflow parameters.  
Table \ref{tbl:inclination} presents the correction factors for a mean 
inclination angle $\langle \theta \rangle \approx 57.3^{\circ}$ 
assuming all orientations have the equally probability 
\citep{1996A&A...311..858B}, nearly pole-on ($5^{\circ}$) and 
nearly edge-on ($85^{\circ}$) inclination angles 
\citep[e.g.,][]{2014ApJ...783...29D}.

For the mean inclination angle the  $t_{\rm dyn}$ decreases by a 
factor of 0.6, while the $P_{\rm out}$, $E_{\rm out}$ and 
$\dot{M}_{\rm out}$ increase by factors of 1.9, 3.4 and 1.7, 
respectively.  The correction factors for $t_{\rm dyn}$ are greater 
and smaller than 1 for $\theta <$ 45$^{\circ}$  and 
$\theta >$ 45$^{\circ}$, respectively. The correction factors for 
$P_{\rm out}$ and $E_{\rm out}$ are always larger than or equal to 
1 for all of possible inclination angle. 
For $\dot{M}_{\rm out}$, its correction factors are  greater 
and smaller than 1 for $\theta >$ 45$^{\circ}$  and 
$\theta <$ 45$^{\circ}$, respectively. 
The possibility of inclination angles $\theta >$ 45$^{\circ}$, 
$\int_{45}^{90} {\rm sin\, \theta\,} d\theta$, is 71\%, which is higher 
than  that of $\theta <$ 45$^{\circ}$, $\int_{0}^{45} {\rm sin\, \theta\,} 
d\theta \sim 29\%$.  This indicates that the inclination correction 
factors for $t_{\rm dyn}$ and $\dot{M}_{\rm out}$, are typically 
smaller and greater than 1, respectively. 
The correlation between $t_{\rm dyn}$ and gas mass of dense 
cores could be affected by the outflow inclination,  while this 
correlation can persist if the outflow inclinations are not in the 
most extreme cases (Figure \ref{fig:outflow-para}).

\subsection{Evolutionary stage}
Infrared emission can be an indicator of star formation 
activities in molecular clouds  since thermal radiation 
from embedded protostars heats the surrounding dust and 
gas. The clumps in this study are dark in the infrared 
wavelengths up to 70~\um, and are believed to be at an 
extreme early phase of star formation, possibly 
encompassing starless and prestellar candidates 
(Figure \ref{fig:cont1}). In total, we have uncovered 
44  dense cores in seven 70~\um\ dark clumps using 
1.3~mm continuum image. Seventeen 
of these dense cores are associated with CO outflows and 
5 of these 17 were identified as having high velocity CO 
components. This indicates that these 70~\um\ dark clumps 
have already begun forming stars, and not all of embedded 
dense cores are starless and/or prestellar 
\citep[see also][]{2019A&A...622A..54P}. There are  
some dense cores associated with very weak or no line 
emission (e.g., CO/$^{13}$CO/C$^{18}$O, CH$_{3}$OH, 
H$_{2}$CO, HCN and HCO$^{+}$), which could be starless 
or prestellar candidates. With the molecular line 
emission we  found that there are chemical differentiation
between different dense cores within the same parent 
clumps, which suggests that they maybe at different 
evolutionary stages. A detailed analysis of spectral 
lines is beyond the scope of this paper, which will be 
presented in a forthcoming publication (Li et al., in 
preparation).

For each clump, the most massive cores  are associated 
with CO outflows, with the exception of AGAL030, which 
shows no outflow signatures toward its embedded dense 
cores. All CO outflows are associated with massive cores 
($>$ 8 \Mo), with the exception of AGAL022-MM1 that 
contains gas mass of 4.1 \Mo. We cannot fully rule out that 
the lack  of CO outflow detections toward the intermediate- 
and low-mass cores  due to the limited sensitivity of our 
observations.  

In addition, we find that 
the outflow dynamical timescale in the massive cores 
tend to be longer than that of less massive cores. In a 
protocluster, if massive protostars are formed at an 
earlier stage than their low-mass counterparts 
\citep[e.g.,][]{2015ApJ...804..141Z}, we expect to see 
outflow activities in the massive cores earlier than the 
less massive ones. On the other hand, the outflow 
dynamical timescale might be affected by sensitivity of 
observations arising from mostly missing flux since 
without single dish data to recover missing short 
spacing fluxes in the SMA data. 
Apart from that, the outflow dynamical timescale is also 
biased  by the outflow inclination 
(see Section \ref{inclination}).

AGAL030 is one exception that its five embedded  dense 
cores are not associated with molecular outflow. 
There are no CO, $^{13}$CO, C$^{18}$O, SiO and HCN 
emission toward these dense cores, while faint HCO$^{+}$
emission is detected toward MM1 and MM2.  In addition, faint 
H$_{2}$CO and CH$_{3}$OH emission are detected toward 
MM1.  We also found CO emission on the west of MM2 
and north-east of MM1, while these emission are produced by 
the ambient gas rather than outflow motions. The observed 
lines emission imply that there are no star-forming signatures 
in these five embedded cores. This indicates that these cores 
could be starless or prestellar candidates.  The averaged-volume 
and surface densities, $\Sigma=M/(\pi r^{2}$),  
of AGAL030 are similar to the AGAL022 that only detected one 
outflow motion, while is relatively lower than other five clumps 
that detected significant outflow motions (see Table \ref{tbl:Jean}). 
The dense cores embedded in higher densities clumps are most 
likely associated with outflow motions for this 70~$\mu$m dark 
cloud sample.  
This speculation can be tested by studying a large sample of 
70~$\mu$m dark cloud.

\section{Summary}
We studied  a sample of seven massive 70~\um\ dark 
clumps in both dust continuum and the CO 2-1 emission 
through high angular resolution observations using the 
SMA. The main findings are as follows.

\begin{itemize}
  \item In total, 44 cores are identified based on the 
  1.3~mm dust continuum emission. They have masses ranging
  from 1.4 up to 77.1 \Mo\ with effective radius of 
  0.02 -- 0.11 pc. Among the 44 cores, 29 dense cores have 
  masses in the range 8.1 to 77.1 \Mo\ and the other 15 
  cores with masses between 1.4 and 7.5 \Mo. 
  Assuming the core density follows  a power-law distribution 
  in radius $\rho \propto r^{-b}$, the 
  power-law index $b$ is between 0.6 and 2.1, while majority
  of dense cores (70\%) have an index between 1 and 2. 

  \item The outflow activities revealed by  CO 
  emission are detected toward 17 dense cores that are 
  embedded in 6 clumps. The CO outflows are only detected 
  toward the massive cores ($>$ 8 \Mo), except for 
  AGAL022-MM1 (4.1 \Mo). Five out of seventeen dense cores are 
  associated with high-velocity CO outflows. These results
  suggest that these 70 \um\ dark clumps host protostars, 
  and therefore they are not quiescent in star formation.  
  
   \item There appears to be a positive correlation between 
   outflow dynamical timescale and gas mass of dense 
   cores. The outflows from massive cores have longer 
   dynamical timescale than that of less massive cores, 
   which indicates that the massive cores have a longer 
   accretion history than the less massive ones. 
   In addition, the outflow dynamical timescales are comparable  
   to the free-fall timescales with a same order of magnitude 
   of $\sim$10$^{4}$~yr.

  \item We find that the gas masses of 
  identified low-mass dense cores are roughly consistent 
  with the pure thermal Jeans fragmentation, while massive 
  cores are approximately consistent with the turbulent Jean 
  fragmentation. The identified cores are not in a state 
  of virial  equilibrium when magnetic fields and external
  cloud pressure are not considered. 
     
\end{itemize}

\acknowledgments
This work is supported by the National Key R\&D Program 
of China (No. 2017YFA0402604). 
SL acknowledges support from the CfA pre-doctoral 
fellowship, the Chinese Scholarship Council and National 
Natural Science Foundation of China grant 11629302 and 
U1731237. 

\vspace{5mm}
\facilities{SMA, APEX, \textit{Herschel} }

\software{MIR, MIRIAD \citep{1995ASPC...77..433S}, 
CASA \citep{2007ASPC..376..127M}, 
APLpy \citep{2012ascl.soft08017R},  
Astropy \citep{2013A&A...558A..33A}, 
LMFIT \citep{newville_matthew_2014_11813}. }

\bibliographystyle{aasjournal}
\bibliography{lsh-IRDCs}

\begin{thebibliography}{}
\expandafter\ifx\csname natexlab\endcsname\relax\def\natexlab#1{#1}\fi

\bibitem[{{Astropy Collaboration} {et~al.}(2013){Astropy Collaboration},
  {Robitaille}, {Tollerud}, {Greenfield}, {Droettboom}, {Bray}, {Aldcroft},
  {Davis}, {Ginsburg}, {Price-Whelan}, {Kerzendorf}, {Conley}, {Crighton},
  {Barbary}, {Muna}, {Ferguson}, {Grollier}, {Parikh}, {Nair}, {Unther},
  {Deil}, {Woillez}, {Conseil}, {Kramer}, {Turner}, {Singer}, {Fox}, {Weaver},
  {Zabalza}, {Edwards}, {Azalee Bostroem}, {Burke}, {Casey}, {Crawford},
  {Dencheva}, {Ely}, {Jenness}, {Labrie}, {Lim}, {Pierfederici}, {Pontzen},
  {Ptak}, {Refsdal}, {Servillat}, \& {Streicher}}]{2013A&A...558A..33A}
{Astropy Collaboration}, {Robitaille}, T.~P., {Tollerud}, E.~J., {et~al.} 2013,
  \aap, 558, A33

\bibitem[{{Ballesteros-Paredes} {et~al.}(2018){Ballesteros-Paredes},
  {V{\'a}zquez-Semadeni}, {Palau}, \& {Klessen}}]{2018MNRAS.479.2112B}
{Ballesteros-Paredes}, J., {V{\'a}zquez-Semadeni}, E., {Palau}, A., \&
  {Klessen}, R.~S. 2018, \mnras, 479, 2112

\bibitem[{{Bally} \& {Lada}(1983)}]{1983ApJ...265..824B}
{Bally}, J., \& {Lada}, C.~J. 1983, \apj, 265, 824

\bibitem[{{Beuther} {et~al.}(2005){Beuther}, {Schilke}, {Menten}, {Motte},
  {Sridharan}, \& {Wyrowski}}]{2005ApJ...633..535B}
{Beuther}, H., {Schilke}, P., {Menten}, K.~M., {et~al.} 2005, \apj, 633, 535

\bibitem[{{Beuther} {et~al.}(2002){Beuther}, {Schilke}, {Sridharan}, {Menten},
  {Walmsley}, \& {Wyrowski}}]{2002A&A...383..892B}
{Beuther}, H., {Schilke}, P., {Sridharan}, T.~K., {et~al.} 2002, \aap, 383, 892

\bibitem[{{Beuther} {et~al.}(2018){Beuther}, {Mottram}, {Ahmadi}, {Bosco},
  {Linz}, {Henning}, {Klaassen}, {Winters}, {Maud}, \&
  {Kuiper}}]{2018A&A...617A.100B}
{Beuther}, H., {Mottram}, J.~C., {Ahmadi}, A., {et~al.} 2018, \aap, 617, A100

\bibitem[{{Blake} {et~al.}(1987){Blake}, {Sutton}, {Masson}, \&
  {Phillips}}]{1987ApJ...315..621B}
{Blake}, G.~A., {Sutton}, E.~C., {Masson}, C.~R., \& {Phillips}, T.~G. 1987,
  \apj, 315, 621

\bibitem[{{Bonnell} \& {Bate}(2002)}]{2002MNRAS.336..659B}
{Bonnell}, I.~A., \& {Bate}, M.~R. 2002, \mnras, 336, 659

\bibitem[{{Bonnell} \& {Bate}(2006)}]{2006MNRAS.370..488B}
---. 2006, \mnras, 370, 488

\bibitem[{{Bonnell} {et~al.}(2001){Bonnell}, {Bate}, {Clarke}, \&
  {Pringle}}]{2001MNRAS.323..785B}
{Bonnell}, I.~A., {Bate}, M.~R., {Clarke}, C.~J., \& {Pringle}, J.~E. 2001,
  \mnras, 323, 785

\bibitem[{{Bontemps} {et~al.}(1996){Bontemps}, {Andre}, {Terebey}, \&
  {Cabrit}}]{1996A&A...311..858B}
{Bontemps}, S., {Andre}, P., {Terebey}, S., \& {Cabrit}, S. 1996, \aap, 311,
  858

\bibitem[{{Butler} \& {Tan}(2012)}]{2012ApJ...754....5B}
{Butler}, M.~J., \& {Tan}, J.~C. 2012, \apj, 754, 5

\bibitem[{{Cabrit} \& {Bertout}(1992)}]{1992A&A...261..274C}
{Cabrit}, S., \& {Bertout}, C. 1992, \aap, 261, 274

\bibitem[{{Carey} {et~al.}(1998){Carey}, {Clark}, {Egan}, {Price}, {Shipman},
  \& {Kuchar}}]{1998ApJ...508..721C}
{Carey}, S.~J., {Clark}, F.~O., {Egan}, M.~P., {et~al.} 1998, \apj, 508, 721

\bibitem[{{Cyganowski} {et~al.}(2017){Cyganowski}, {Brogan}, {Hunter}, {Smith},
  {Kruijssen}, {Bonnell}, \& {Zhang}}]{2017MNRAS.468.3694C}
{Cyganowski}, C.~J., {Brogan}, C.~L., {Hunter}, T.~R., {et~al.} 2017, \mnras,
  468, 3694

\bibitem[{{Devereux} \& {Young}(1990)}]{1990ApJ...359...42D}
{Devereux}, N.~A., \& {Young}, J.~S. 1990, \apj, 359, 42

\bibitem[{{Dunham} {et~al.}(2014){Dunham}, {Arce}, {Mardones}, {Lee},
  {Matthews}, {Stutz}, \& {Williams}}]{2014ApJ...783...29D}
{Dunham}, M.~M., {Arce}, H.~G., {Mardones}, D., {et~al.} 2014, \apj, 783, 29

\bibitem[{{Dunham} {et~al.}(2008){Dunham}, {Crapsi}, {Evans}, {Bourke},
  {Huard}, {Myers}, \& {Kauffmann}}]{2008ApJS..179..249D}
{Dunham}, M.~M., {Crapsi}, A., {Evans}, Neal~J., I., {et~al.} 2008, \apjs, 179,
  249

\bibitem[{{Elia} {et~al.}(2017){Elia}, {Molinari}, {Schisano}, {Pestalozzi},
  {Pezzuto}, {Merello}, {Noriega-Crespo}, {Moore}, {Russeil}, {Mottram},
  {Paladini}, {Strafella}, {Benedettini}, {Bernard}, {Di Giorgio}, {Eden},
  {Fukui}, {Plume}, {Bally}, {Martin}, {Ragan}, {Jaffa}, {Motte}, {Olmi},
  {Schneider}, {Testi}, {Wyrowski}, {Zavagno}, {Calzoletti}, {Faustini},
  {Natoli}, {Palmeirim}, {Piacentini}, {Piazzo}, {Pilbratt}, {Polychroni},
  {Baldeschi}, {Beltr{\'a}n}, {Billot}, {Cambr{\'e}sy}, {Cesaroni},
  {Garc{\'\i}a-Lario}, {Hoare}, {Huang}, {Joncas}, {Liu}, {Maiolo}, {Marsh},
  {Maruccia}, {M{\`e}ge}, {Peretto}, {Rygl}, {Schilke}, {Thompson},
  {Traficante}, {Umana}, {Veneziani}, {Ward-Thompson}, {Whitworth}, {Arab},
  {Band ieramonte}, {Becciani}, {Brescia}, {Buemi}, {Bufano}, {Butora},
  {Cavuoti}, {Costa}, {Fiorellino}, {Hajnal}, {Hayakawa}, {Kacsuk}, {Leto}, {Li
  Causi}, {Marchili}, {Martinavarro-Armengol}, {Mercurio}, {Molinaro},
  {Riccio}, {Sano}, {Sciacca}, {Tachihara}, {Torii}, {Trigilio}, {Vitello}, \&
  {Yamamoto}}]{2017MNRAS.471..100E}
{Elia}, D., {Molinari}, S., {Schisano}, E., {et~al.} 2017, \mnras, 471, 100

\bibitem[{{Foster} {et~al.}(2009){Foster}, {Rosolowsky}, {Kauffmann}, {Pineda},
  {Borkin}, {Caselli}, {Myers}, \& {Goodman}}]{2009ApJ...696..298F}
{Foster}, J.~B., {Rosolowsky}, E.~W., {Kauffmann}, J., {et~al.} 2009, \apj,
  696, 298

\bibitem[{{Goldsmith}(2001)}]{2001ApJ...557..736G}
{Goldsmith}, P.~F. 2001, The Astrophysical Journal, 557, 736

\bibitem[{{Hacar} {et~al.}(2016){Hacar}, {Alves}, {Burkert}, \&
  {Goldsmith}}]{2016A&A...591A.104H}
{Hacar}, A., {Alves}, J., {Burkert}, A., \& {Goldsmith}, P. 2016, Astronomy and
  Astrophysics, 591, A104

\bibitem[{{Henning} {et~al.}(2010){Henning}, {Linz}, {Krause}, {Ragan},
  {Beuther}, {Launhardt}, {Nielbock}, \& {Vasyunina}}]{2010A&A...518L..95H}
{Henning}, T., {Linz}, H., {Krause}, O., {et~al.} 2010, \aap, 518, L95

\bibitem[{{Henshaw} {et~al.}(2017){Henshaw}, {Jim{\'e}nez-Serra}, {Longmore},
  {Caselli}, {Pineda}, {Avison}, {Barnes}, {Tan}, \&
  {Fontani}}]{2017MNRAS.464L..31H}
{Henshaw}, J.~D., {Jim{\'e}nez-Serra}, I., {Longmore}, S.~N., {et~al.} 2017,
  \mnras, 464, L31

\bibitem[{{Hildebrand}(1983)}]{1983QJRAS..24..267H}
{Hildebrand}, R.~H. 1983, \qjras, 24, 267

\bibitem[{{Ho} {et~al.}(2004){Ho}, {Moran}, \& {Lo}}]{2004ApJ...616L...1H}
{Ho}, P.~T.~P., {Moran}, J.~M., \& {Lo}, K.~Y. 2004, \apjl, 616, L1

\bibitem[{{Hull} \& {Zhang}(2019)}]{2019FrASS...6....3H}
{Hull}, C. L.~H., \& {Zhang}, Q. 2019, Frontiers in Astronomy and Space
  Sciences, 6, 3

\bibitem[{{Kauffmann} {et~al.}(2008){Kauffmann}, {Bertoldi}, {Bourke}, {Evans},
  \& {Lee}}]{2008A&A...487..993K}
{Kauffmann}, J., {Bertoldi}, F., {Bourke}, T.~L., {Evans}, II, N.~J., \& {Lee},
  C.~W. 2008, \aap, 487, 993

\bibitem[{{Kauffmann} {et~al.}(2013){Kauffmann}, {Pillai}, \&
  {Goldsmith}}]{2013ApJ...779..185K}
{Kauffmann}, J., {Pillai}, T., \& {Goldsmith}, P.~F. 2013, \apj, 779, 185

\bibitem[{{Keto}(2003)}]{2003ApJ...599.1196K}
{Keto}, E. 2003, \apj, 599, 1196

\bibitem[{{Kippenhahn} {et~al.}(2012){Kippenhahn}, {Weigert}, \&
  {Weiss}}]{2012sse..book.....K}
{Kippenhahn}, R., {Weigert}, A., \& {Weiss}, A. 2012, {Stellar Structure and
  Evolution}, doi:10.1007/978-3-642-30304-3

\bibitem[{{Kirk} {et~al.}(2017){Kirk}, {Friesen}, {Pineda}, {Rosolowsky},
  {Offner}, {Matzner}, {Myers}, {Di Francesco}, {Caselli}, {Alves},
  {Chac{\'o}n-Tanarro}, {Chen}, {Chun-Yuan Chen}, {Keown}, {Punanova}, {Seo},
  {Shirley}, {Ginsburg}, {Hall}, {Singh}, {Arce}, {Goodman}, {Martin}, \&
  {Redaelli}}]{2017ApJ...846..144K}
{Kirk}, H., {Friesen}, R.~K., {Pineda}, J.~E., {et~al.} 2017, \apj, 846, 144

\bibitem[{{Krumholz} {et~al.}(2007){Krumholz}, {Klein}, \&
  {McKee}}]{2007ApJ...656..959K}
{Krumholz}, M.~R., {Klein}, R.~I., \& {McKee}, C.~F. 2007, \apj, 656, 959

\bibitem[{{Krumholz} {et~al.}(2005){Krumholz}, {McKee}, \&
  {Klein}}]{2005Natur.438..332K}
{Krumholz}, M.~R., {McKee}, C.~F., \& {Klein}, R.~I. 2005, \nat, 438, 332

\bibitem[{{Lada} {et~al.}(2008){Lada}, {Muench}, {Rathborne}, {Alves}, \&
  {Lombardi}}]{2008ApJ...672..410L}
{Lada}, C.~J., {Muench}, A.~A., {Rathborne}, J., {Alves}, J.~F., \& {Lombardi},
  M. 2008, \apj, 672, 410

\bibitem[{{Looney} {et~al.}(2000){Looney}, {Mundy}, \&
  {Welch}}]{2000ApJ...529..477L}
{Looney}, L.~W., {Mundy}, L.~G., \& {Welch}, W.~J. 2000, \apj, 529, 477

\bibitem[{{Lu} {et~al.}(2015){Lu}, {Zhang}, {Wang}, \&
  {Gu}}]{2015ApJ...805..171L}
{Lu}, X., {Zhang}, Q., {Wang}, K., \& {Gu}, Q. 2015, \apj, 805, 171

\bibitem[{{MacLaren} {et~al.}(1988){MacLaren}, {Richardson}, \&
  {Wolfendale}}]{1988ApJ...333..821M}
{MacLaren}, I., {Richardson}, K.~M., \& {Wolfendale}, A.~W. 1988, \apj, 333,
  821

\bibitem[{{Mangum} \& {Shirley}(2015)}]{2015PASP..127..266M}
{Mangum}, J.~G., \& {Shirley}, Y.~L. 2015, \pasp, 127, 266

\bibitem[{{McKee} \& {Tan}(2002)}]{2002Natur.416...59M}
{McKee}, C.~F., \& {Tan}, J.~C. 2002, \nat, 416, 59

\bibitem[{{McKee} \& {Tan}(2003)}]{2003ApJ...585..850M}
---. 2003, \apj, 585, 850

\bibitem[{{McMullin} {et~al.}(2007){McMullin}, {Waters}, {Schiebel}, {Young},
  \& {Golap}}]{2007ASPC..376..127M}
{McMullin}, J.~P., {Waters}, B., {Schiebel}, D., {Young}, W., \& {Golap}, K.
  2007, in Astronomical Society of the Pacific Conference Series, Vol. 376,
  Astronomical Data Analysis Software and Systems XVI, ed. R.~A. {Shaw},
  F.~{Hill}, \& D.~J. {Bell}, 127

\bibitem[{{Motte} \& {Andr{\'e}}(2001)}]{2001A&A...365..440M}
{Motte}, F., \& {Andr{\'e}}, P. 2001, \aap, 365, 440

\bibitem[{{Motte} {et~al.}(2018){Motte}, {Bontemps}, \&
  {Louvet}}]{2018ARA&A..56...41M}
{Motte}, F., {Bontemps}, S., \& {Louvet}, F. 2018, \araa, 56, 41

\bibitem[{{Mueller} {et~al.}(2002){Mueller}, {Shirley}, {Evans}, \&
  {Jacobson}}]{2002ApJS..143..469M}
{Mueller}, K.~E., {Shirley}, Y.~L., {Evans}, II, N.~J., \& {Jacobson}, H.~R.
  2002, \apjs, 143, 469

\bibitem[{{Myers}(1983)}]{1983ApJ...270..105M}
{Myers}, P.~C. 1983, The Astrophysical Journal, 270, 105

\bibitem[{Newville {et~al.}(2014)Newville, Stensitzki, Allen, \&
  Ingargiola}]{newville_matthew_2014_11813}
Newville, M., Stensitzki, T., Allen, D.~B., \& Ingargiola, A. 2014, {LMFIT:
  Non-Linear Least-Square Minimization and Curve-Fitting for Python}, , ,
  doi:10.5281/zenodo.11813

\bibitem[{{Palau} {et~al.}(2013){Palau}, {Fuente}, {Girart}, {Estalella}, {Ho},
  {S{\'a}nchez-Monge}, {Fontani}, {Busquet}, {Commer{\c c}on}, {Hennebelle},
  {Boissier}, {Zhang}, {Cesaroni}, \& {Zapata}}]{2013ApJ...762..120P}
{Palau}, A., {Fuente}, A., {Girart}, J.~M., {et~al.} 2013, \apj, 762, 120

\bibitem[{{Palau} {et~al.}(2014){Palau}, {Estalella}, {Girart}, {Fuente},
  {Fontani}, {Commer{\c{c}}on}, {Busquet}, {Bontemps}, {S{\'a}nchez-Monge}, \&
  {Zapata}}]{2014ApJ...785...42P}
{Palau}, A., {Estalella}, R., {Girart}, J.~M., {et~al.} 2014, \apj, 785, 42

\bibitem[{{Palau} {et~al.}(2015){Palau}, {Ballesteros-Paredes},
  {V{\'a}zquez-Semadeni}, {S{\'a}nchez-Monge}, {Estalella}, {Fall}, {Zapata},
  {Camacho}, {G{\'o}mez}, {Naranjo-Romero}, {Busquet}, \&
  {Fontani}}]{2015MNRAS.453.3785P}
{Palau}, A., {Ballesteros-Paredes}, J., {V{\'a}zquez-Semadeni}, E., {et~al.}
  2015, \mnras, 453, 3785

\bibitem[{{Palau} {et~al.}(2018){Palau}, {Zapata}, {Rom{\'a}n-Z{\'u}{\~n}iga},
  {S{\'a}nchez-Monge}, {Estalella}, {Busquet}, {Girart}, {Fuente}, \&
  {Commer{\c c}on}}]{2018ApJ...855...24P}
{Palau}, A., {Zapata}, L.~A., {Rom{\'a}n-Z{\'u}{\~n}iga}, C.~G., {et~al.} 2018,
  \apj, 855, 24

\bibitem[{{Pattle} {et~al.}(2015){Pattle}, {Ward-Thompson}, {Kirk}, {White},
  {Drabek-Maunder}, {Buckle}, {Beaulieu}, {Berry}, {Broekhoven-Fiene},
  {Currie}, {Fich}, {Hatchell}, {Kirk}, {Jenness}, {Johnstone}, {Mottram},
  {Nutter}, {Pineda}, {Quinn}, {Salji}, {Tisi}, {Walker-Smith}, {di Francesco},
  {Hogerheijde}, {Andr{\'e}}, {Bastien}, {Bresnahan}, {Butner}, {Chen},
  {Chrysostomou}, {Coude}, {Davis}, {Duarte-Cabral}, {Fiege}, {Friberg},
  {Friesen}, {Fuller}, {Graves}, {Greaves}, {Gregson}, {Griffin}, {Holland},
  {Joncas}, {Knee}, {K{\"o}nyves}, {Mairs}, {Marsh}, {Matthews},
  {Moriarty-Schieven}, {Rawlings}, {Richer}, {Robertson}, {Rosolowsky},
  {Rumble}, {Sadavoy}, {Spinoglio}, {Thomas}, {Tothill}, {Viti}, {Wouterloot},
  {Yates}, \& {Zhu}}]{2015MNRAS.450.1094P}
{Pattle}, K., {Ward-Thompson}, D., {Kirk}, J.~M., {et~al.} 2015, \mnras, 450,
  1094

\bibitem[{{Pillai} {et~al.}(2011){Pillai}, {Kauffmann}, {Wyrowski}, {Hatchell},
  {Gibb}, \& {Thompson}}]{2011A&A...530A.118P}
{Pillai}, T., {Kauffmann}, J., {Wyrowski}, F., {et~al.} 2011, \aap, 530, A118

\bibitem[{{Pillai} {et~al.}(2019){Pillai}, {Kauffmann}, {Zhang}, {Sanhueza},
  {Leurini}, {Wang}, {Sridharan}, \& {K{\"o}nig}}]{2019A&A...622A..54P}
{Pillai}, T., {Kauffmann}, J., {Zhang}, Q., {et~al.} 2019, \aap, 622, A54

\bibitem[{{Pokhrel} {et~al.}(2018){Pokhrel}, {Myers}, {Dunham}, {Stephens},
  {Sadavoy}, {Zhang}, {Bourke}, {Tobin}, {Lee}, {Gutermuth}, \&
  {Offner}}]{2018ApJ...853....5P}
{Pokhrel}, R., {Myers}, P.~C., {Dunham}, M.~M., {et~al.} 2018, \apj, 853, 5

\bibitem[{{Qiu} {et~al.}(2009){Qiu}, {Zhang}, {Wu}, \&
  {Chen}}]{2009ApJ...696...66Q}
{Qiu}, K., {Zhang}, Q., {Wu}, J., \& {Chen}, H.-R. 2009, \apj, 696, 66

\bibitem[{{Ragan} {et~al.}(2012){Ragan}, {Henning}, {Krause}, {Pitann},
  {Beuther}, {Linz}, {Tackenberg}, {Balog}, {Hennemann}, {Launhardt}, {Lippok},
  {Nielbock}, {Schmiedeke}, {Schuller}, {Steinacker}, {Stutz}, \&
  {Vasyunina}}]{2012A&A...547A..49R}
{Ragan}, S., {Henning}, T., {Krause}, O., {et~al.} 2012, \aap, 547, A49

\bibitem[{{Rathborne} {et~al.}(2006){Rathborne}, {Jackson}, \&
  {Simon}}]{2006ApJ...641..389R}
{Rathborne}, J.~M., {Jackson}, J.~M., \& {Simon}, R. 2006, \apj, 641, 389

\bibitem[{{Reid} {et~al.}(2009){Reid}, {Menten}, {Zheng}, {Brunthaler},
  {Moscadelli}, {Xu}, {Zhang}, {Sato}, {Honma}, {Hirota}, {Hachisuka}, {Choi},
  {Moellenbrock}, \& {Bartkiewicz}}]{2009ApJ...700..137R}
{Reid}, M.~J., {Menten}, K.~M., {Zheng}, X.~W., {et~al.} 2009, \apj, 700, 137

\bibitem[{{Richer} {et~al.}(2000){Richer}, {Shepherd}, {Cabrit}, {Bachiller},
  \& {Churchwell}}]{2000prpl.conf..867R}
{Richer}, J.~S., {Shepherd}, D.~S., {Cabrit}, S., {Bachiller}, R., \&
  {Churchwell}, E. 2000, Protostars and Planets IV, 867

\bibitem[{{Robitaille} \& {Bressert}(2012)}]{2012ascl.soft08017R}
{Robitaille}, T., \& {Bressert}, E. 2012, {APLpy: Astronomical Plotting Library
  in Python}, Astrophysics Source Code Library, , , ascl:1208.017

\bibitem[{{Rosolowsky} {et~al.}(2008){Rosolowsky}, {Pineda}, {Kauffmann}, \&
  {Goodman}}]{2008ApJ...679.1338R}
{Rosolowsky}, E.~W., {Pineda}, J.~E., {Kauffmann}, J., \& {Goodman}, A.~A.
  2008, \apj, 679, 1338

\bibitem[{{Sanhueza} {et~al.}(2017){Sanhueza}, {Jackson}, {Zhang},
  {Guzm{\'a}n}, {Lu}, {Stephens}, {Wang}, \& {Tatematsu}}]{2017ApJ...841...97S}
{Sanhueza}, P., {Jackson}, J.~M., {Zhang}, Q., {et~al.} 2017, \apj, 841, 97

\bibitem[{{Sault} {et~al.}(1995){Sault}, {Teuben}, \&
  {Wright}}]{1995ASPC...77..433S}
{Sault}, R.~J., {Teuben}, P.~J., \& {Wright}, M.~C.~H. 1995, in Astronomical
  Society of the Pacific Conference Series, Vol.~77, Astronomical Data Analysis
  Software and Systems IV, ed. R.~A. {Shaw}, H.~E. {Payne}, \& J.~J.~E.
  {Hayes}, 433

\bibitem[{{Schuller} {et~al.}(2009){Schuller}, {Menten}, {Contreras},
  {Wyrowski}, {Schilke}, {Bronfman}, {Henning}, {Walmsley}, {Beuther},
  {Bontemps}, {Cesaroni}, {Deharveng}, {Garay}, {Herpin}, {Lefloch}, {Linz},
  {Mardones}, {Minier}, {Molinari}, {Motte}, {Nyman}, {Reveret}, {Risacher},
  {Russeil}, {Schneider}, {Testi}, {Troost}, {Vasyunina}, {Wienen}, {Zavagno},
  {Kovacs}, {Kreysa}, {Siringo}, \& {Wei{\ss}}}]{2009A&A...504..415S}
{Schuller}, F., {Menten}, K.~M., {Contreras}, Y., {et~al.} 2009, \aap, 504, 415

\bibitem[{{Scoville} \& {Kwan}(1976)}]{1976ApJ...206..718S}
{Scoville}, N.~Z., \& {Kwan}, J. 1976, \apj, 206, 718

\bibitem[{{Shirley} {et~al.}(2002){Shirley}, {Evans}, \&
  {Rawlings}}]{2002ApJ...575..337S}
{Shirley}, Y.~L., {Evans}, Neal~J., I., \& {Rawlings}, J. M.~C. 2002, \apj,
  575, 337

\bibitem[{{Shu} {et~al.}(1987){Shu}, {Adams}, \&
  {Lizano}}]{1987ARA&A..25...23S}
{Shu}, F.~H., {Adams}, F.~C., \& {Lizano}, S. 1987, \araa, 25, 23

\bibitem[{{Shu} {et~al.}(2000){Shu}, {Najita}, {Shang}, \&
  {Li}}]{2000prpl.conf..789S}
{Shu}, F.~H., {Najita}, J.~R., {Shang}, H., \& {Li}, Z.-Y. 2000, Protostars and
  Planets IV, 789

\bibitem[{{Stephens} {et~al.}(2018){Stephens}, {Dunham}, {Myers}, {Pokhrel},
  {Bourke}, {Vorobyov}, {Tobin}, {Sadavoy}, {Pineda}, {Offner}, {Lee},
  {Kristensen}, {J{\o}rgensen}, {Goodman}, {Arce}, \&
  {Gurwell}}]{2018ApJS..237...22S}
{Stephens}, I.~W., {Dunham}, M.~M., {Myers}, P.~C., {et~al.} 2018, The
  Astrophysical Journal Supplement Series, 237, 22

\bibitem[{{Stutz} {et~al.}(2013){Stutz}, {Tobin}, {Stanke}, {Megeath},
  {Fischer}, {Robitaille}, {Henning}, {Ali}, {di Francesco}, {Furlan},
  {Hartmann}, {Osorio}, {Wilson}, {Allen}, {Krause}, \&
  {Manoj}}]{2013ApJ...767...36S}
{Stutz}, A.~M., {Tobin}, J.~J., {Stanke}, T., {et~al.} 2013, \apj, 767, 36

\bibitem[{{Tobin} {et~al.}(2015){Tobin}, {Stutz}, {Megeath}, {Fischer},
  {Henning}, {Ragan}, {Ali}, {Stanke}, {Manoj}, {Calvet}, \&
  {Hartmann}}]{2015ApJ...798..128T}
{Tobin}, J.~J., {Stutz}, A.~M., {Megeath}, S.~T., {et~al.} 2015, \apj, 798, 128

\bibitem[{{Tomisaka}(1998)}]{1998ApJ...502L.163T}
{Tomisaka}, K. 1998, \apjl, 502, L163

\bibitem[{{Traficante} {et~al.}(2018){Traficante}, {Lee}, {Hennebelle},
  {Molinari}, {Kauffmann}, \& {Pillai}}]{2018A&A...619L...7T}
{Traficante}, A., {Lee}, Y.-N., {Hennebelle}, P., {et~al.} 2018, \aap, 619, L7

\bibitem[{{Urquhart} {et~al.}(2014){Urquhart}, {Moore}, {Csengeri}, {Wyrowski},
  {Schuller}, {Hoare}, {Lumsden}, {Mottram}, {Thompson}, {Menten}, {Walmsley},
  {Bronfman}, {Pfalzner}, {K{\"o}nig}, \& {Wienen}}]{2014MNRAS.443.1555U}
{Urquhart}, J.~S., {Moore}, T.~J.~T., {Csengeri}, T., {et~al.} 2014, \mnras,
  443, 1555

\bibitem[{{Urquhart} {et~al.}(2015){Urquhart}, {Figura}, {Moore}, {Csengeri},
  {Lumsden}, {Pillai}, {Thompson}, {Eden}, \& {Morgan}}]{2015MNRAS.452.4029U}
{Urquhart}, J.~S., {Figura}, C.~C., {Moore}, T.~J.~T., {et~al.} 2015, \mnras,
  452, 4029

\bibitem[{{van der Tak} {et~al.}(2000){van der Tak}, {van Dishoeck}, {Evans},
  \& {Blake}}]{2000ApJ...537..283V}
{van der Tak}, F.~F.~S., {van Dishoeck}, E.~F., {Evans}, II, N.~J., \& {Blake},
  G.~A. 2000, \apj, 537, 283

\bibitem[{{Vuong} {et~al.}(2003){Vuong}, {Montmerle}, {Grosso}, {Feigelson},
  {Verstraete}, \& {Ozawa}}]{2003A&A...408..581V}
{Vuong}, M.~H., {Montmerle}, T., {Grosso}, N., {et~al.} 2003, \aap, 408, 581

\bibitem[{{Wang} {et~al.}(2012){Wang}, {Zhang}, {Wu}, {Li}, \&
  {Zhang}}]{2012ApJ...745L..30W}
{Wang}, K., {Zhang}, Q., {Wu}, Y., {Li}, H.-b., \& {Zhang}, H. 2012, \apj, 745,
  L30

\bibitem[{{Wang} {et~al.}(2011){Wang}, {Zhang}, {Wu}, \&
  {Zhang}}]{2011ApJ...735...64W}
{Wang}, K., {Zhang}, Q., {Wu}, Y., \& {Zhang}, H. 2011, \apj, 735, 64

\bibitem[{{Wang} {et~al.}(2014){Wang}, {Zhang}, {Testi}, {van der Tak}, {Wu},
  {Zhang}, {Pillai}, {Wyrowski}, {Carey}, {Ragan}, \&
  {Henning}}]{2014MNRAS.439.3275W}
{Wang}, K., {Zhang}, Q., {Testi}, L., {et~al.} 2014, \mnras, 439, 3275

\bibitem[{{Wang} {et~al.}(2008){Wang}, {Zhang}, {Pillai}, {Wyrowski}, \&
  {Wu}}]{2008ApJ...672L..33W}
{Wang}, Y., {Zhang}, Q., {Pillai}, T., {Wyrowski}, F., \& {Wu}, Y. 2008, \apjl,
  672, L33

\bibitem[{{Wang} {et~al.}(2006){Wang}, {Zhang}, {Rathborne}, {Jackson}, \&
  {Wu}}]{2006ApJ...651L.125W}
{Wang}, Y., {Zhang}, Q., {Rathborne}, J.~M., {Jackson}, J., \& {Wu}, Y. 2006,
  \apjl, 651, L125

\bibitem[{{Wienen} {et~al.}(2012){Wienen}, {Wyrowski}, {Schuller}, {Menten},
  {Walmsley}, {Bronfman}, \& {Motte}}]{2012A&A...544A.146W}
{Wienen}, M., {Wyrowski}, F., {Schuller}, F., {et~al.} 2012, \aap, 544, A146

\bibitem[{{Zhang} {et~al.}(2005){Zhang}, {Hunter}, {Brand}, {Sridharan},
  {Cesaroni}, {Molinari}, {Wang}, \& {Kramer}}]{2005ApJ...625..864Z}
{Zhang}, Q., {Hunter}, T.~R., {Brand}, J., {et~al.} 2005, \apj, 625, 864

\bibitem[{{Zhang} \& {Wang}(2011)}]{2011ApJ...733...26Z}
{Zhang}, Q., \& {Wang}, K. 2011, \apj, 733, 26

\bibitem[{{Zhang} {et~al.}(2015){Zhang}, {Wang}, {Lu}, \&
  {Jim{\'e}nez-Serra}}]{2015ApJ...804..141Z}
{Zhang}, Q., {Wang}, K., {Lu}, X., \& {Jim{\'e}nez-Serra}, I. 2015, \apj, 804,
  141

\bibitem[{{Zhang} {et~al.}(2009){Zhang}, {Wang}, {Pillai}, \&
  {Rathborne}}]{2009ApJ...696..268Z}
{Zhang}, Q., {Wang}, Y., {Pillai}, T., \& {Rathborne}, J. 2009, \apj, 696, 268

\bibitem[{{Zinnecker} \& {Yorke}(2007)}]{2007ARA&A..45..481Z}
{Zinnecker}, H., \& {Yorke}, H.~W. 2007, Annual Review of Astronomy and
  Astrophysics, 45, 481

\end{thebibliography}

%

\begin{figure*}[ht!]
\epsscale{1.1}
\plotone{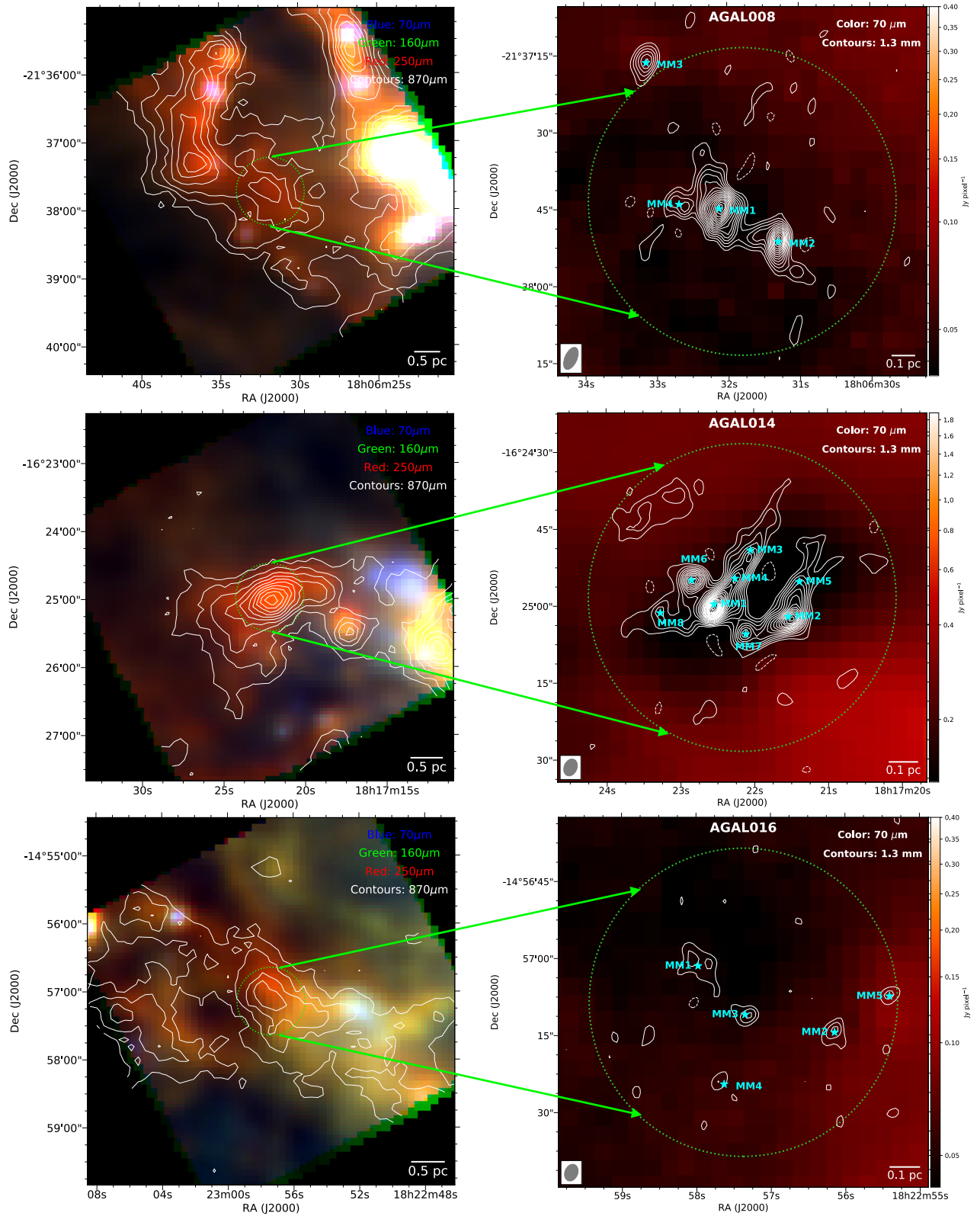}
\caption{Left: the APEX 870 $\mu$m emission 
(white contours) overlaid on three-color \textit{Herschel} 
composite image (blue/green/red = 70/160/250 $\mu$m). 
The white contours are $\pm$(4, 8, 12 ....)$\times$ $\sigma$, 
where the $\sigma$ is the rms level for each source.
Right: the 1.3 mm dust continuum overlaid on  
\textit{Herschel} 70 $\mu$m. The white contours are
$\pm$(3, 5, 7, 9 ....)$\times$ $\sigma$. The dash circle is 
the SMA primary beam FWHM size. The synthesized beam 
are shown in the bottom left corner of the image. The dash 
green circle is the SMA primary beam FWHM size.}
\label{fig:cont1}
\end{figure*}

\begin{figure*}[ht!]\ContinuedFloat 
\epsscale{1.1}
\plotone{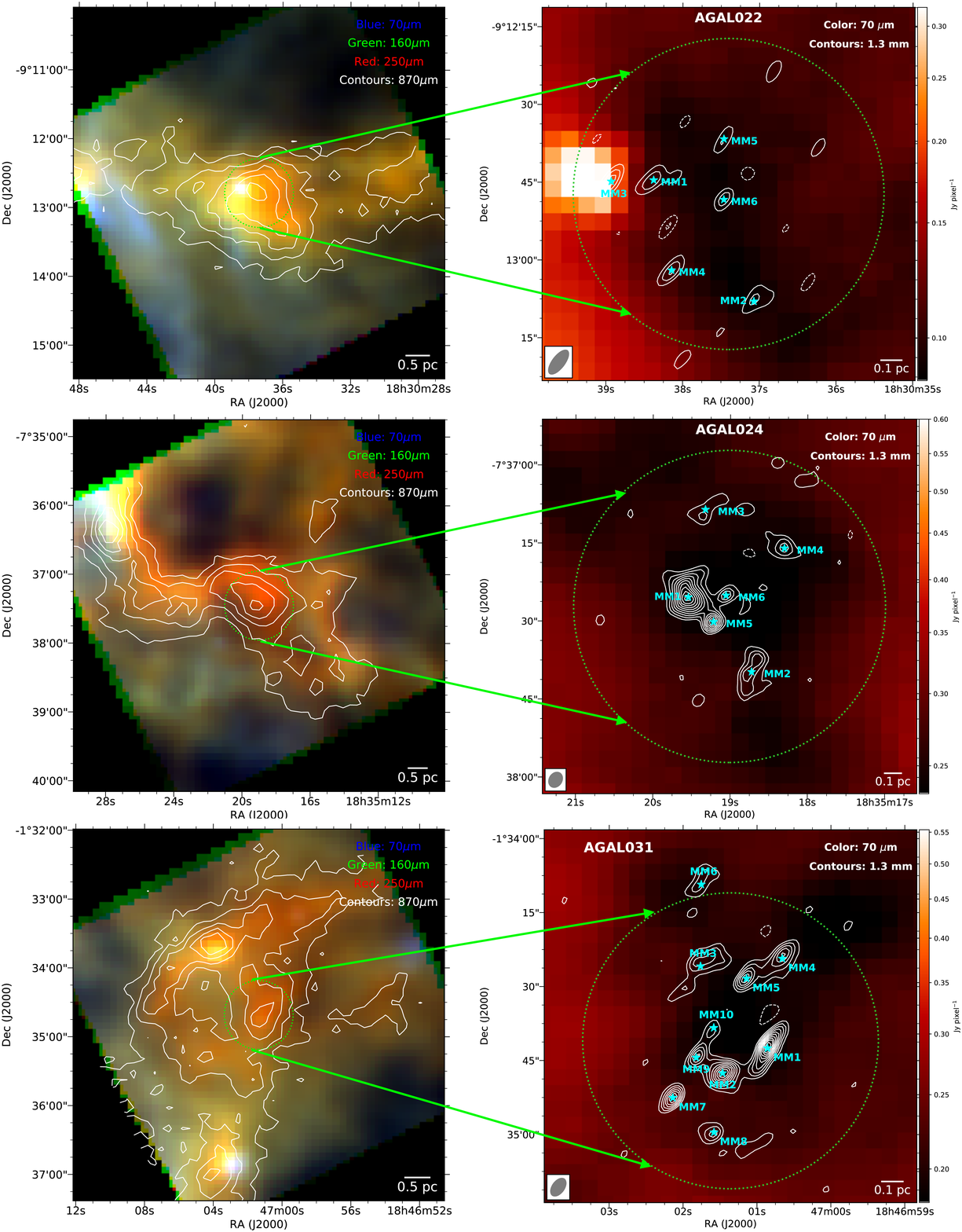}
\caption{Continuation}
\end{figure*}  

\begin{figure*}[ht!]\ContinuedFloat 
\epsscale{1.1}
\plotone{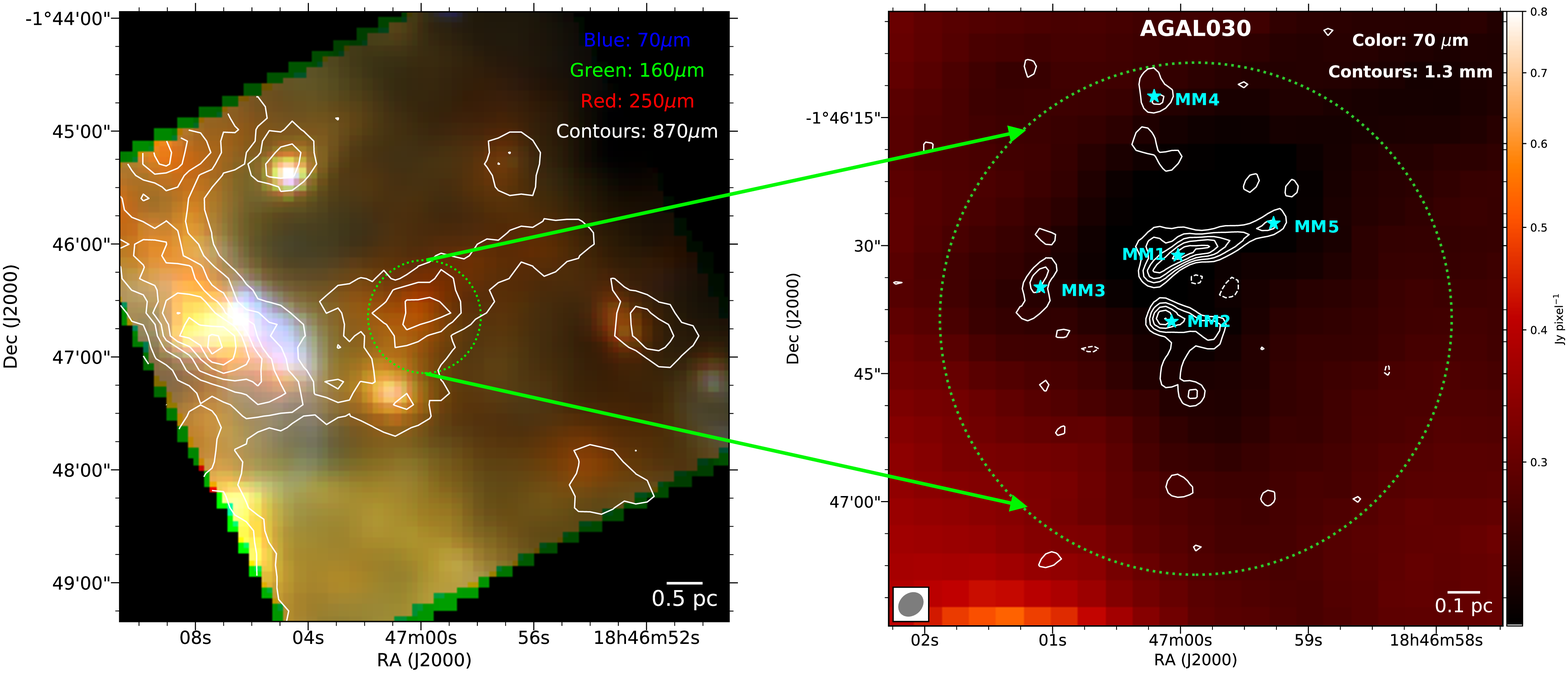}
\caption{Continuation}
\end{figure*}

\floattable
\begin{deluxetable*}{clcccccccccccccc}
\tabletypesize{\scriptsize}
\tablecolumns{13}
\tablewidth{0pc}
\tablecaption{Observations. \label{tbl:obs}}
\tablehead{
\colhead{Date} &\colhead{Sources}&\colhead{R.A.}&\colhead{Decl.} &\colhead{$D$}&\colhead{Calibrator}  &\colhead{Calibrator} 
&\colhead{Calibrator} &\colhead{$N_{\rm ant}$} \\
  &&\colhead{(J2000)}&\colhead{(J2000)} &\colhead{(kpc)}&\colhead{(Gain)}&\colhead{(Flux)}&\colhead{(Bandpass)}  &}
\startdata
\multirow{2}{*}{08/Jul/2017}
&AGAL030.844+00.177&18:46:59.88&-01:46:38.6 &5.4&1751+096, 1830+063    &Ganymede&3c279  & 7\\
&AGAL024.314+00.086&18:35:19.01& -07:37:27.2   &5.9&1751+096, 1830+063    &Ganymede&3c279  & 7\\
          \hline    
\multirow{2}{*}{25/Jul/2017}
&AGAL014.492-00.139&18:17:22.16& -16:24:58.4 &3.6&nrao530, 1924-292    &Neptune&3c279  & 8\\
&AGAL016.418-00.634&18:22:57.37& -14:57:08.6&3.5&nrao530, 1924-292    &Neptune&3c279  &  8\\
          \hline      
\multirow{2}{*}{01/Aug/2017}
&AGAL031.024+00.262&18:47:01.36&-01:34:41.0&4.5&1743-038, 1830+063&Neptune&3c454.3  & 7 \\
&AGAL008.691-00.401&18:06:31.80&-21:37:43.6&4.5&1743-038, 1830+063&Neptune&3c454.3  & 7 \\
          \hline    
\multirow{1}{*}{06/Aug/2017}
&AGAL022.376+00.447&18:30:37.40&-09:12:47.2&3.8&1743-038, 1830+063    &Neptune&3c454.3  & 6 \\
\enddata
\tablenotetext{}{Notes. Columns are (1)observational day, (2) source name, (3) right ascension, (4) declination, 
(5)target distance, (6)gain calibrator, (7)flux calibrator, (8) bandpass calibrator, (9) number of antenna.  \\ }
\end{deluxetable*}

\floattable
\begin{deluxetable*}{clccccccccccccccccccc}
\tabletypesize{\scriptsize}
\rotate
\tablecolumns{17}
\tablewidth{0pc}
\tablecaption{Dense core physical parameters. 
\label{tbl:cores}
}
\tablehead{
\colhead{Sources}&\colhead{Core ID}&\colhead{R.A.}&\colhead{Decl.} &\colhead{$\sigma_{maj}$} &\colhead{$\sigma_{min}$}&\colhead{PA } &\colhead{$F_{\nu}$}&\colhead{$S_{\nu}^{\rm beam}$}&\colhead{$M_{\rm gas}$}&\colhead{$N_{\rm H_{2}}$} &\colhead{$n_{\rm H_{2}}$} &\colhead{$R_{\rm eff}$}  &\colhead{$R_{\rm eff(dec)}$} &\colhead{$\sigma_{\rm obs}$}   &\colhead{$M_{\rm vir}$}    &\colhead{$\alpha_{\rm vir}$}  &\colhead{$b$}  &\colhead{$T_{\rm K}$}&\colhead{CO}&\colhead{SiO} \\
 &&\colhead{(hh:mm:ss)}&\colhead{(dd:mm:ss)} &\colhead{(arcsec)}  &\colhead{(arcsec)}    &\colhead{(deg)}    &\colhead{(mJy)}   &\colhead{(mJy beam$^{-1}$)}  &\colhead{(M$_{\odot}$)}
&\colhead{(cm$^{-2}$)} &\colhead{(cm$^{-3}$)} &\colhead{(pc)}  &\colhead{(pc)}  &\colhead{(km s$^{-1}$)} &\colhead{(M$_{\odot}$)}  & & & \colhead{(K)} &\colhead{} &\colhead{}
}
\startdata
          \hline  
\multirow{4}{*}{AGLA008}
&MM1&18:06:32.13&-21:37:45.00&8.1 &5.6 &25 &7.43E-02&1.78E-02&55.78 &1.08E+23&5.59E+05&0.09 &0.07 &1.00 &69.40 &1.24 &1.2$\pm$0.1&16 &Y&Y\\
&MM2&18:06:31.30 &-21:37:51.48&6.3 &5.3 &21 &4.22E-02&1.38E-02&31.69 &8.35E+22&6.46E+05&0.08 &0.06 &0.91 &23.85 &0.75 &2.1$\pm$0.1&16 &Y&N\\
&MM3&18:06:33.13&-21:37:16.43&5.5 &3.5 &163 &4.94E-02&2.77E-02&37.13 &1.68E+23&2.78E+06&0.06 &0.04 &...&...&...&...&16 &Y&N\\
&MM4&18:06:32.68&-21:37:44.11&5.1 &4.2 &58 &1.08E-02&5.49E-03&8.14 &3.33E+22&...&0.06 &...&...&...&...&...&16 &N&N\\
          \hline 
\multirow{10}{*}{AGAL014}
&MM1&18:17:22.54&-16:24:59.70&5.6 &5.5 &77 &1.14E-01&3.17E-02&47.29 &1.98E+23&1.41E+06&0.058 &0.049 &...&...&...&1.3 $\pm$0.1 &16 &Y&N\\
&MM2&18:17:21.54&-16:25:02.16&8.3 &4.4 &141 &1.02E-01&2.39E-02&42.18 &1.49E+23&9.00E+05&0.063 &0.054 &...&...&...&1.1 $\pm$0.1 &16 &Y&Y\\
&MM3&18:17:22.05&-16:24:49.19&9.9 &6.6 &38 &4.47E-02&5.88E-03&18.49 &3.67E+22&1.38E+05&0.085 &0.077 &1.00 &75.21 &4.07 &...&16 &N&N\\
&MM4&18:17:22.26&-16:24:54.68&7.0 &4.2 &128 &3.31E-02&9.69E-03&13.71 &6.04E+22&4.81E+05&0.057 &0.046 &0.74 &14.89 &1.09 &...&16 &N&N\\
&MM5&18:17:21.39&-16:24:55.35&8.5 &6.4 &102 &3.32E-02&5.29E-03&13.75 &3.30E+22&1.41E+05&0.077 &0.069 &0.79 &27.34 &1.99 &...&16 &N&N\\
&MM6&18:17:22.85&-16:24:55.00&4.6 &3.7 &133 &3.13E-02&1.59E-02&12.97 &9.91E+22&1.78E+06&0.043 &0.029 &...&...&...&1.4 $\pm$0.2 &16 &Y&N\\
&MM7&18:17:22.12&-16:25:05.54&4.4 &4.3 &136 &2.62E-02&1.17E-02&10.83 &7.29E+22&1.04E+06&0.046 &0.033 &0.99 &34.05 &3.14 &0.9 $\pm$0.3 &16 &Y&Y\\
&MM8&18:17:23.27&-16:25:01.20&8.1 &6.0 &129 &3.05E-02&5.37E-03&12.65 &3.35E+22&1.51E+05&0.073 &0.066 &...&...&...&...&16 &N&N\\
           \hline 
\multirow{5}{*}{AGAL016}
&MM1&18:22:57.98&-14:57:01.49&9.3 &6.2 &23 &2.18E-02&3.21E-03&10.95 &2.65E+22&1.11E+05&0.08 &0.07 &...&...&...&...&13 &Y&N\\
&MM2&18:22:56.15&-14:57:14.47&4.8 &3.9 &179 &1.21E-02&5.48E-03&6.07 &4.52E+22&6.52E+05&0.04 &0.03 &...&...&...&...&13 &Y&Y\\
&MM3&18:22:57.35&-14:57:10.98&4.8 &3.5 &68 &6.66E-03&3.37E-03&3.35 &2.78E+22&1.43E+06&0.04 &0.02 &...&...&...&...&13 &N&N\\
&MM4&18:22:57.63&-14:57:24.53&5.7 &4.1 &36 &8.88E-03&3.20E-03&4.46 &2.64E+22&3.16E+05&0.05 &0.04 &...&...&...&...&13 &N&N\\
&MM5&18:22:55.41&-14:57:07.47&4.7 &3.5 &150 &1.29E-02&6.73E-03&6.49 &5.55E+22&1.01E+06&0.04 &0.03 &...&...&...&...&13 &N&N\\
          \hline  
\multirow{6}{*}{AGAL022}
&MM1&18:30:38.38&-9:12:44.57&7.9 &3.5 &117 &8.91E-03&4.39E-03&4.10 &1.75E+22&...&0.06 &...&...&...&...&...&16 &Y&N\\
&MM2&18:30:37.07&-9:13:07.95&8.4 &3.5 &129 &1.05E-02&4.72E-03&4.82 &1.88E+22&2.49E+05&0.06 &0.04 &...&...&...&...&16 &N&N\\
&MM3&18:30:38.93&-9:12:44.83&6.4 &2.7 &142 &9.26E-03&7.28E-03&4.26 &2.90E+22&1.56E+06&0.05 &0.02 &0.696&5.45 &1.28 &...&16 &N&N\\
&MM4&18:30:38.14&-9:13:01.96&6.9 &2.6 &138 &6.97E-03&5.14E-03&3.20 &2.04E+22&1.06E+06&0.05 &0.02 &...&...&...&...&16 &N&N\\
&MM5&18:30:37.46&-9:12:36.64&6.7 &2.4 &151 &3.48E-03&2.88E-03&1.60 &1.14E+22&...&0.04 &...&0.454&...&...&...&16 &N&N\\
&MM6&18:30:37.47&-9:12:48.32&4.4 &3.0 &143 &3.02E-03&3.12E-03&1.39 &1.24E+22&...&0.04 &...&...&...&...&...&16 &N&N\\
           \hline 
\multirow{6}{*}{AGAL024}
&MM1&18:35:19.54 &-7:37:25.39&7.1 &5.3 &18 &5.71E-02&1.19E-02&77.06 &9.78E+22&3.34E+05&0.11 &0.09 &0.71 &24.87 &0.32 &1.3$\pm$0.2&14&Y&Y\\
&MM2&18:35:18.72 &-7:37:39.74&8.9 &4.1 &164 &2.54E-02&5.50E-03&34.28 &4.52E+22&1.70E+05&0.10 &0.09 &0.62 &15.54 &0.45 &...&14&Y&N\\
&MM3&18:35:19.32 &-7:37:08.54&11.5 &4.2 &97 &2.80E-02&4.51E-03&37.81 &3.71E+22&1.27E+05&0.12 &0.10 &0.57 &13.44 &0.36 &...&14&N&N\\
&MM4&18:35:18.30 &-7:37:15.97&6.7 &3.3 &58 &1.55E-02&5.46E-03&20.94 &4.49E+22&7.90E+05&0.08 &0.05 &...&...&...&...&14&Y&N\\
&MM5&18:35:19.22 &-7:37:30.12&3.2 &2.6 &152 &8.88E-03&8.24E-03&11.99 &6.78E+22&2.20E+07&0.05 &0.01 &0.75 &2.88 &0.24 &2$\pm$0.1&14&N&N\\
&MM6&18:35:19.06 &-7:37:25.04&4.0 &2.3 &127 &5.53E-03&4.61E-03&7.47 &3.79E+22&...&0.05 &...&...&...&...&...&14&N&N\\
           \hline 
\multirow{10}{*}{AGAL031}
&MM1&18:47:00.86&-1:34:42.49&8.3 &2.9 &148 &2.42E-02&1.04E-02&20.59 &7.02E+22&9.54E+05&0.064 &0.042 &...&...&...&1.8$\pm$0.1&13 &Y&Y\\
&MM2&18:47:01.47&-1:34:47.55&5.3 &4.3 &56 &2.13E-02&9.66E-03&18.18 &6.54E+22&2.12E+06&0.063 &0.031 &0.76 &11.21 &0.62 &1$\pm$0.1&13 &Y&N\\
&MM3&18:47:01.76&-1:34:25.88&10.2 &4.5 &135 &2.34E-02&5.20E-03&19.96 &3.52E+22&1.44E+05&0.090 &0.078 &...&...&...&...&13 &N&N\\
&MM4&18:47:00.66&-1:34:24.31&7.4 &3.7 &142 &2.01E-02&7.57E-03&17.12 &5.12E+22&3.79E+05&0.069 &0.054 &0.78 &21.94 &1.28 &0.9$\pm$0.2&13 &Y&N\\
&MM5&18:47:01.14&-1:34:28.36&7.0 &3.0 &136 &1.52E-02&7.33E-03&12.99 &4.96E+22&6.47E+05&0.061 &0.041 &...&...&...&0.6$\pm$0.3&13 &Y&N\\
&MM6&18:47:01.76&-1:34:09.29&10.0 &4.5 &153 &4.05E-02&9.22E-03&34.56 &6.24E+22&2.62E+05&0.089 &0.077 &...&...&...&...&13 &N&N\\
&MM7&18:47:02.14&-1:34:52.55&4.5 &2.6 &143 &1.19E-02&1.03E-02&10.14 &6.95E+22&7.52E+06&0.045 &0.017 &0.84 &7.69 &0.76 &1.5$\pm$0.3&13 &Y&N\\
&MM8&18:47:01.58&-1:34:59.55&4.6 &4.1 &7 &1.08E-02&5.85E-03&9.17 &3.96E+22&1.42E+06&0.057 &0.028 &0.60 &4.27 &0.46 &...&13 &N&N\\
&MM9&18:47:01.82&-1:34:44.39&5.4 &2.2 &149 &6.79E-03&5.75E-03&5.79 &3.89E+22&...&0.046 &...&0.61 &...&...&...&13 &N&N\\
&MM10&18:47:01.58&-1:34:38.35&6.7 &3.2 &151 &4.71E-03&2.20E-03&4.01 &1.49E+22&1.66E+05&0.062 &...&...&...&...&...&13 &N&N\\
          \hline
\multirow{5}{*}{AGAL030}
&MM1&18:47:00.02&-1:46:31.15&8.5 &2.6 &110 &1.79E-02&5.98E-03&20.21 &5.12E+22&1.22E+06&0.07 &0.04 &...&...&...&...&14&N&N\\
&MM2&18:47:00.06&-1:46:38.71&4.4 &2.6 &60 &6.00E-03&3.92E-03&6.76 &3.35E+22&...&0.05 &...&...&...&...&...&14&N&N\\
&MM3&18:47:01.09&-1:46:34.88&5.8 &2.6 &162 &8.79E-03&4.28E-03&9.90 &3.67E+22&8.38E+06&0.06 &0.02 &...&...&...&...&14&N&N\\
&MM4&18:47:00.21&-1:46:12.50&4.5 &3.6 &4 &1.30E-02&6.00E-03&14.69 &5.14E+22&5.90E+05&0.06 &0.04 &...&...&...&...&14&N&N\\
&MM5&18:46:59.27&-1:46:27.38&5.5 &4.0 &94 &6.61E-03&2.25E-03&7.44 &1.93E+22&1.26E+05&0.07 &0.06 &...&...&...&...&14&N&N\\
\enddata
\tablenotetext{}{Notes. Sources: source name. Core ID: dense core.  R.A.: right ascension. Decl.: declination. $\sigma_{maj}$: beam-convolved major axis. $\sigma_{min}$: beam-convolved minor axis. PA: position angle.  $F_{\nu}$: total integrated flux. $S_{\nu}^{\rm beam}$: peak flux density. $M_{\rm gas}$: gas mass. $N_{\rm H_{2}}$: column density. $n_{\rm H_{2}}$: volumn density. $R_{\rm eff}$: beam-convolved effective radius.  $R_{\rm eff(dec)}$: beam-deconvolved effective radius. $\sigma_{\rm obs}$: channel width deconvolved line width of C$^{18}$O.   $M_{\rm vir}$: virial mass. $\alpha_{\rm vir}$: virial parameters. \new{$b$: best fitted density power-law index.} $T_{\rm K}$: kinematical temperature. CO: CO outflow detection. SiO: SiO detection.}
\end{deluxetable*}

\floattable
\begin{deluxetable*}{clccccccccccccccc}
\tabletypesize{\scriptsize}
\tablecolumns{13}
\tablewidth{0pc}
\tablecaption{Outflow parameters. 
\label{tbl:outflow}}
\tablehead{
\colhead{Source} &\colhead{Core ID} &&\colhead{$\Delta v$}&\colhead{$M_{\rm out}$}&\colhead{$P_{\rm out}$}             
&\colhead{$E_{\rm out}$}    &\colhead{$t_{\rm dyn}$}                     &\colhead{$\dot{M}_{\rm out}$}                    
& \colhead{$l_{\rm out}$}\\
 &&    &\colhead{(km s$^{-1}$)}&\colhead{(M$_{\odot}$)}&\colhead{(M$_{\odot}$ km s$^{-1}$)} 
&\colhead{(M$_{\odot}$ km$^{2}$ s$^{-2}$)}&\colhead{(10$^{4}$ yr)}   &\colhead{(10$^{-5}$ M$_{\odot}$ yr$^{-1}$)}    
&\colhead{(pc)}}
\startdata
\multirow{6}{*}{AGAL008}
&\multirow{2}{*}{MM1}
&Blue& [19, 35] & 0.035 &0.281 &1.36 &2.61 &0.136 &0.51 \\
&&Red& [41, 70] & 0.031 &0.272 &1.43 &0.48 &0.647 &0.15 \\
\cline{2-10}
&\multirow{2}{*}{MM2}
&Blue& [30, 38] & 0.016 &0.080 &0.22 &0.81 &0.203 &0.07 \\
&&Red& [39, 48] & 0.010 &0.043 &0.10 &1.67 &0.058 &0.15 \\
\cline{2-10}
&\multirow{2}{*}{MM3}
&Blue& [21, 35] & 0.012 &0.096 &0.50 &0.95 &0.13 &0.17 \\
&&Red& [43, 61] & 0.011 &0.112 &0.63 &0.73 &0.15 &0.17 \\
\hline
\multirow{6}{*}{AGAL014}
&{MM1}&Red& [41, 56] & 0.061 &0.353 &1.49 &2.51 &0.24 &0.41 \\
\cline{2-10}
&\multirow{2}{*}{MM2}
&Blue& [3, 38] & 0.017 &0.250 &2.43 &0.35 &0.49 &0.13 \\
&&Red& [40, 96] & 0.148 &2.42 &32.36 &0.41 &3.59 &0.23 \\
\cline{2-10}
&\multirow{2}{*}{MM6}
&Blue& [15, 38] & 0.048 &0.466 &2.94 &0.94 &0.52 &0.24 \\
&&Red& [43, 75] & 0.036 &0.432 &3.45 &0.52 &0.70 &0.18 \\
\cline{2-10}
&MM7&Blue& [11, 35] & 0.024 &0.306 &2.09 &0.40 &0.60 &0.12 \\
\hline
\multirow{2}{*}{AGAL016}
&\multirow{1}{*}{MM1}
&Red& [46, 58] & 0.006 &0.045 &0.19 &0.9 &0.06 &0.15 \\
\cline{2-10}
&\multirow{1}{*}{MM2}
&Red& [46, 59] & 0.003 &0.032 &0.16 &0.7 &0.05 &0.11 \\
\cline{2-10}
\hline
\multirow{1}{*}{AGAL022}
&\multirow{1}{*}{MM1}
&Red& [83, 89] & 0.01 &0.04 &0.09 &1.55 & 0.06 &0.11 \\
\hline
\multirow{4}{*}{AGAL024}
&\multirow{2}{*}{MM1}
&Blue& [96, 112] & 0.006 &0.037 &0.15 &1.61 &0.04 &0.28 \\
&&Red& [117, 133] & 0.087 &0.827 &4.31 &1.62 &0.54 &0.33 \\
\cline{2-10}
&{MM2}&Red& [120, 133] & 0.050 &0.403 &1.74 &2.69 &0.19 &0.48 \\
\cline{2-10}
&MM4&Red& [117, 134] & 0.055 &0.279 &0.83 &2.52 &0.22 &0.41 \\
\hline
\multirow{9}{*}{AGAL031}
&\multirow{2}{*}{MM1}
&Blue& [49, 78] & 0.036 &0.965 &13.82 &0.38 &0.55 &0.18 \\
&&Red& [97, 141] & 0.087 &1.225 &12.18 &0.55 &0.93 &0.24 \\
\cline{2-10}
&MM2&Red& [96, 114] & 0.113 &0.851 &4.30 &1.22 &2.57 &0.22 \\
\cline{2-10}
&\multirow{2}{*}{MM4}
&Blue& [72, 78] & 0.007 &0.15 &1.65 &0.42 &0.17 &0.11 \\
&&Red& [97, 113] & 0.051 &0.33 &1.42 &1.04 &0.50 &0.17 \\
\cline{2-10}
&\multirow{2}{*}{MM5}
&Blue& [73, 78] & 0.019 &0.387 &3.99 &1.02 &0.18 &0.25 \\
&&Red& [97, 141] & 0.043 &0.544 &5.19 &0.32 &0.50 &0.15 \\
\cline{2-10}
&\multirow{2}{*}{MM7}
&Blue& [52, 77] & 0.050 &1.171 &14.02 &0.47 &1.08 &0.21 \\
&&Red& [83, 94] & 0.028  &0.320 &1.93 &0.78 &0.35 &0.13 \\
\hline
\enddata
\tablenotetext{}{Notes. $\Delta v$: velocity range for blue- and red-shifted components. $M_{\rm out}$: outflow mass. $P_{\rm out}$: outflow momentum. $E_{\rm out}$: outflow energy. $t_{\rm dyn}$: outflow dynamical time. $\dot{M}_{\rm out}$: outflow rate. $L_{\rm out}$: outflow physical length.}
\end{deluxetable*}

\floattable
\begin{deluxetable*}{clcccccccccccccc}
\tabletypesize{\scriptsize}
\tablecolumns{13}
\tablewidth{0pc}
\tablecaption{Jean analysis in the clump. \label{tbl:Jean}}
\tablehead{
\colhead{Sources}&\colhead{$M_{\rm clump}$}&\colhead{$R_{\rm eff(dec)}$} &\colhead{$M_{\rm J}^{\rm th}$}&\colhead{$M_{\rm J}^{\rm tur}$}&\colhead{$L_{\rm J}^{\rm th}$} &\colhead{$L_{\rm J}^{\rm tur}$}&\colhead{$N_{\rm J}^{\rm th}$} & \colhead{$N_{\rm J}^{\rm tur}$} &\colhead{$N_{\rm core}$} &\colhead{$M_{\rm core}$} 
&\colhead{$L_{\rm core}$}  &\colhead{$\Sigma$} &\colhead{$n_{\rm H_{2}}$} \\
&\colhead{(M$_{\odot}$)}&\colhead{(pc)} &\colhead{(M$_{\odot}$)}&\colhead{(M$_{\odot}$)}&\colhead{(pc)}&\colhead{(pc)}&&&&\colhead{(M$_{\odot}$)}
&\colhead{(pc)}&\colhead{($\rm cm^{-2}$)}&\colhead{($\rm cm^{-3}$)} 
}  
\startdata
AGAL008 & 486  & 0.32  & 2.0  & 32.9  & 0.10  & 0.32  & 243  & 15  & 4 & 8.1-55.8 & 0.18-0.62 & 1.71E+22 & 5.21E+04 \\
AGAL014 & 856  & 0.33  & 1.8  & 113.1  & 0.08  & 0.41  & 476  & 8  & 8 & 10.8-47.3 & 0.11-0.15 & 2.85E+22 & 8.48E+04 \\
AGAL016 & 538  & 0.42  & 2.5  & 56.5  & 0.14  & 0.48  & 215  & 10  & 5 & 3.4-11.0 & 0.22-0.24 & 1.08E+22 & 2.49E+04 \\
AGAL022 & 714  & 0.53  & 4.1  & 66.9  & 0.19  & 0.59  & 173  & 11  & 6 & 1.4-4.8 & 0.21-0.32 & 8.85E+21 & 1.61E+04 \\
AGAL024 & 1181  & 0.54  & 2.7  & ... & 0.14  & ... & 437  & ... & 6 & 7.5-77.1 & 0.16-0.49 & 1.44E+22 & 2.59E+04 \\
AGAL031 & 559  & 0.35  & 1.9  & 20.4  & 0.10  & 0.29  & 294  & 27  & 10 & 4.0-34.6 & 0.14-0.37 & 1.58E+22 & 4.35E+04 \\
AGAL030 & 484  & 0.45  & 3.2  & ... & 0.17  & ... & 151  & ... & 5 & 5.3-20.2 & 0.20-0.50 & 8.46E+21 & 1.82E+04 \\
\enddata
\tablenotetext{}{Notes. $M_{\rm clump}$: total mass of compact clump. $R_{\rm eff(dec)}$: beam-deconvolved effective radius of compact clump. $M_{\rm J}^{\rm th}$: thermal Jeans mass. $M_{\rm J}^{\rm tur}$: turbulent Jeans mass. $L_{\rm J}^{\rm th}$: thermal Jeans length. $L_{\rm J}^{\rm tur}$: turbulent Jeans length. $N_{\rm J}^{\rm th}$: thermal Jean number. $N_{\rm J}^{\rm tur}$: turbulent Jean number. $N_{\rm core}$ number of detected cores. $M_{\rm core}$: observed mass of detected cores. $L_{\rm core}$: observed separation between  nearest neighbor cores. $\Sigma$: the clump surface density. $n_{\rm H_{2}}$: the clump averaged-volumn density. 
\\ }
\end{deluxetable*}

\floattable
\begin{deluxetable*}{ccccc}
\tabletypesize{\scriptsize}
\tablecolumns{5}
\tablewidth{0pc}
\tablecaption{Inclination correction factors. \label{tbl:inclination}}
\tablehead{
\colhead{Outflow parameters}&\colhead{Inclination Dependence}
& &\colhead{Correction angles} & \\
\cline{3-5}
&
&\colhead{$\langle \theta \rangle \approx 57.3^{\circ}$}
&\colhead{$\theta = 5^{\circ}$}
&\colhead{$\theta = 85^{\circ}$}  
}
\startdata
$v_{\rm out}$  &  1/cos$\,\theta$     &  1.9 & 1.0   &  11.5 \\
$l_{\rm out}$   &   1/sin$\,\theta$        &  1.2 & 11.5 &  1.0   \\ 
$t_{\rm dyn}$  &  cos$\,\theta$/sin$\,\theta$ & 0.6 & 11.4 &  0.09  \\ 
$P_{\rm out}$ &   1/cos$\,\theta$    & 1.9 & 1.0  &  11.5   \\ 
$E_{\rm out}$ &   1/cos$^{2}\,\theta$            & 3.4 &  1.1 &  131.6 \\ 
$\dot{M}_{\rm out}$& sin$\,\theta$/cos$\,\theta$& 1.7 & 0.09 & 11.4\\  
\enddata
\end{deluxetable*}



\end{document}